\documentclass[aps,preprint,floats]{revtex4dan}

\usepackage{graphicx}

\def\lsim{\mathrel{\raise.3ex\hbox{$<$\kern-.75em\lower1ex\hbox{$\sim$}}}}
\def\gsim{\mathrel{\raise.3ex\hbox{$>$\kern-.75em\lower1ex\hbox{$\sim$}}}}

\begin{document}

\title {High-energy Neutrino Astronomy:  The Cosmic Ray Connection}
\author{Francis Halzen and Dan Hooper}
\affiliation{
Department of Physics, University of Wisconsin,
1150 University Avenue,
Madison, WI 53706}

\begin{abstract}
This is a review of neutrino astronomy anchored to the observational 
fact that Nature accelerates protons and photons to energies in excess 
of $10^{20}$ and $10^{13}$\,eV, respectively.

Although the discovery of cosmic rays dates back close to a century, we 
do not know how and where they are accelerated. There is evidence that 
the highest energy cosmic rays are extra-galactic --- they cannot be 
contained by our galaxy's magnetic field anyway because their gyroradius 
far exceeds its dimension. Elementary elementary-particle physics 
dictates a universal upper limit on their energy of 
$5\times10^{19}$\,eV, the so-called Greisen-Kuzmin-Zatsepin cutoff; 
however, particles in excess of this energy have been observed by all 
experiments, adding one more puzzle to the cosmic ray mystery. Mystery 
is fertile ground for progress: we will review the facts as well as the 
speculations about the sources.

There is a realistic hope that the oldest problem in astronomy will be 
resolved soon by ambitious experimentation: air shower arrays of 
$10^4$\,km$^2$ area, arrays of air Cerenkov detectors and, the subject 
of this review, kilometer-scale neutrino observatories.

We will review why cosmic accelerators are also expected to be cosmic 
beam dumps producing associated high-energy photon and neutrino beams. 
We will work in detail through an example of a cosmic beam dump, gamma 
ray bursts. These are expected to produce neutrinos from MeV to EeV 
energy by a variety of mechanisms. We will also discuss active galaxies 
and GUT-scale remnants, two other classes of sources speculated to be 
associated with the highest energy cosmic rays. Gamma ray bursts and 
active galaxies are also the sources of the highest energy gamma rays, 
with emission observed up to 20\,TeV, possibly higher.

The important conclusion is that, independently of the specific 
blueprint of the source, it takes a kilometer-scale neutrino observatory 
to detect the neutrino beam associated with the highest energy cosmic 
rays and gamma rays. We also briefly review the ongoing efforts to 
commission such instrumentation.
\end{abstract}

\maketitle
\thispagestyle{empty}

\tableofcontents        

\pagebreak              

\section{The Highest Energy Particles: Cosmic Rays, Photons and
Neutrinos}

\subsection{The New Astronomy}

Conventional astronomy spans 60 octaves in photon frequency, from
$10^4$\,cm radio-waves to $10^{-14}$\,cm gamma rays of GeV energy; see
Fig.\,1. This is an amazing expansion of the power of our eyes which
scan the sky over less than a single octave just above $10^{-5}$\,cm
wavelength. This new astronomy probes the
Universe with new wavelengths, smaller than $10^{-14}$\,cm, or photon
energies larger than 10\,GeV. Besides the traditional signals of
astronomy, gamma rays, gravitational waves, neutrinos and very
high-energy
protons become astronomical messengers from the
Universe. As exemplified time and again, the development of novel
ways of looking into space invariably results in the discovery of
unanticipated phenomena. As is the case with new accelerators,
observing only the predicted will be slightly disappointing.

\begin{figure}[thb]
\centering\leavevmode
\includegraphics[width=5in]{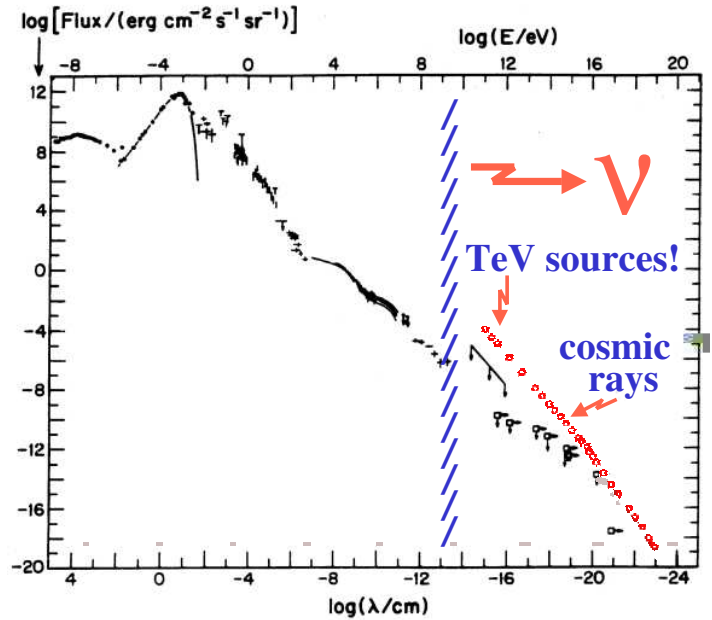}
\caption{The diffuse flux of photons in the Universe, from radio waves to
GeV-photons. Above tens of GeV, only limits are reported although
individual sources emitting TeV gamma rays have been identified. Above
GeV
energy, cosmic rays dominate the spectrum.}
\label{one}
\end{figure}

Why pursue high-energy astronomy with neutrinos or protons despite
considerable instrumental challenges? A mundane
reason is that the
Universe is not transparent to photons of TeV energy and above
(units are: GeV/TeV/PeV/EeV/ZeV in ascending factors of $10^3$). For
instance, a PeV energy photon cannot deliver information from a
source
at the edge of our own galaxy because it will annihilate into an
electron pair in an encounter with a 2.7 Kelvin microwave
photon before reaching our telescope. In general,
energetic
photons are absorbed on background light by pair production $\gamma +
\gamma_{\rm \,bkgnd} \rightarrow e^+ + e^-$ of electrons above a
threshold $E$ given by
\begin{equation}
4E\epsilon \sim (2m_e)^2 \,,
\end{equation}
where $E$ and $\epsilon$ are the energy of the high-energy and
background photon, respectively. Eq.\,(1) implies that TeV-photons are
absorbed
on infrared light, PeV photons on the cosmic microwave background and
EeV photons on radio-waves; see Fig.\,1. Only neutrinos can reach us
without
attenuation from the edge of the Universe.

At EeV energies, proton astronomy may be possible. Near 50\,EeV and
above, the arrival directions of electrically charged cosmic rays are
no longer scrambled by the ambient magnetic field of our own galaxy.
They point back to their sources with an accuracy determined by their
gyroradius in the intergalactic magnetic field $B$:
\begin{equation}
\theta \cong {d\over R_{\rm gyro}} = {dB\over E} \,,
\end{equation}
where $d$ is the distance to the source. Scaled to units relevant to
the problem,
\begin{equation}
{\theta\over0.1^\circ} \cong { \left( d\over 1{\rm\ Mpc} \right)
\left( B\over 10^{-9}{\rm\,G} \right) \over \left( E\over
3\times10^{20}\rm\, eV\right) }\,.
\end{equation}
Speculations on the strength of the inter-galactic magnetic field
range from $10^{-7}$ to $10^{-12}$~Gauss in the local cluster. For a
distance of 100~Mpc,
the resolution may therefore be anywhere from sub-degree to
nonexistent. It is still possible that the arrival
directions of the highest energy cosmic rays provide
information on the location of their sources. Proton astronomy should
be possible; it may also provide indirect information on
intergalactic magnetic fields. Determining the strength of intergalactic
magnetic fields by conventional astronomical means has been challenging.

\subsection{The Highest Energy Cosmic Rays: Facts}

In October 1991, the Fly's Eye cosmic ray detector recorded an event
of energy $3.0\pm^{0.36}_{0.54}\times 10^{20}$\,eV \cite{flyes}.
This event, together with an event recorded by the Yakutsk air shower
array in May 1989 \cite{yakutsk}, of estimated energy $\sim$
$2\times10^{20}$\,eV, constituted (at the time) the two highest
energy cosmic rays ever seen. Their energy corresponds to a center of
mass energy of the order of 700~TeV or $\sim 50$ Joules, almost 50
times the energy of the Large Hadron Collider (LHC). In fact, all active
experiments \cite{web} have detected
cosmic rays in the vicinity of 100~EeV since their initial discovery by
the
Haverah Park air shower array \cite{WatsonZas}. The AGASA air shower
array in Japan\cite{agasa} has now accumulated an impressive 10
events with energy in excess of $10^{20}$\,eV \cite{ICRC}.

\begin{figure}[t!]
\centering\leavevmode
\includegraphics[width=5in]{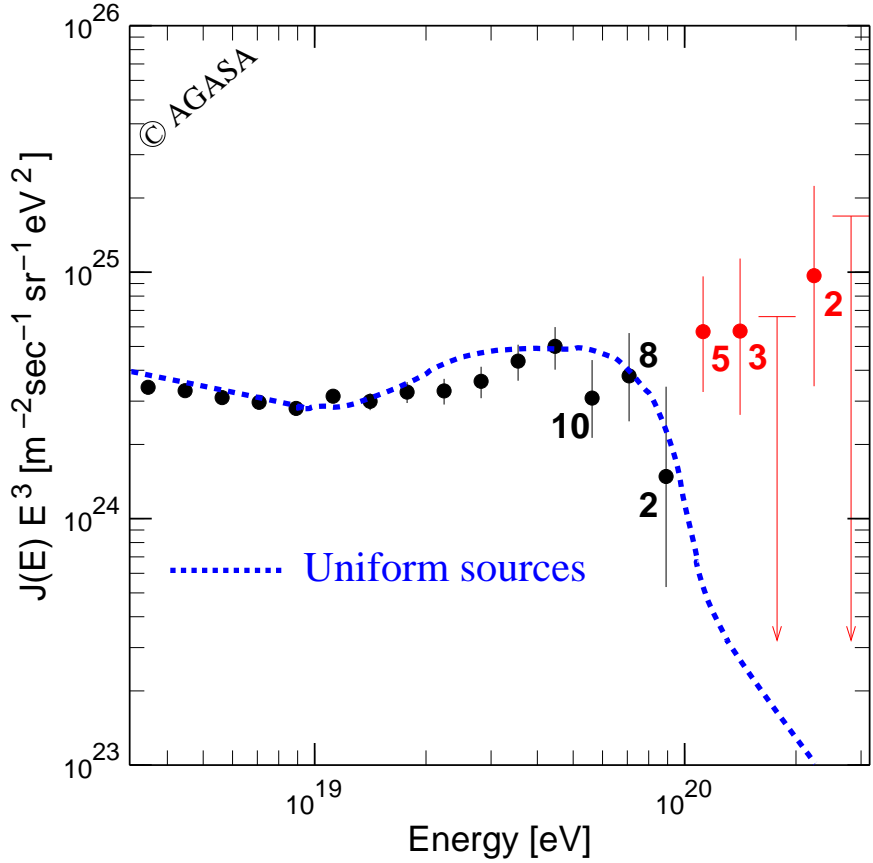}
\caption{The cosmic ray spectrum peaks in the vicinity of 1\,GeV and has
features near $10^{15}$ and $10^{19}$\,eV referred to as the ``knee" and
``ankle" in the spectrum, respectively. Shown is the flux of the highest
energy cosmic rays near and beyond the ankle measured by the AGASA
experiment.  Note that the flux is multiplied by $E^3$.}
\label{two}
\end{figure}

\begin{figure}[t!]
\centering\leavevmode
\includegraphics[width=5in]{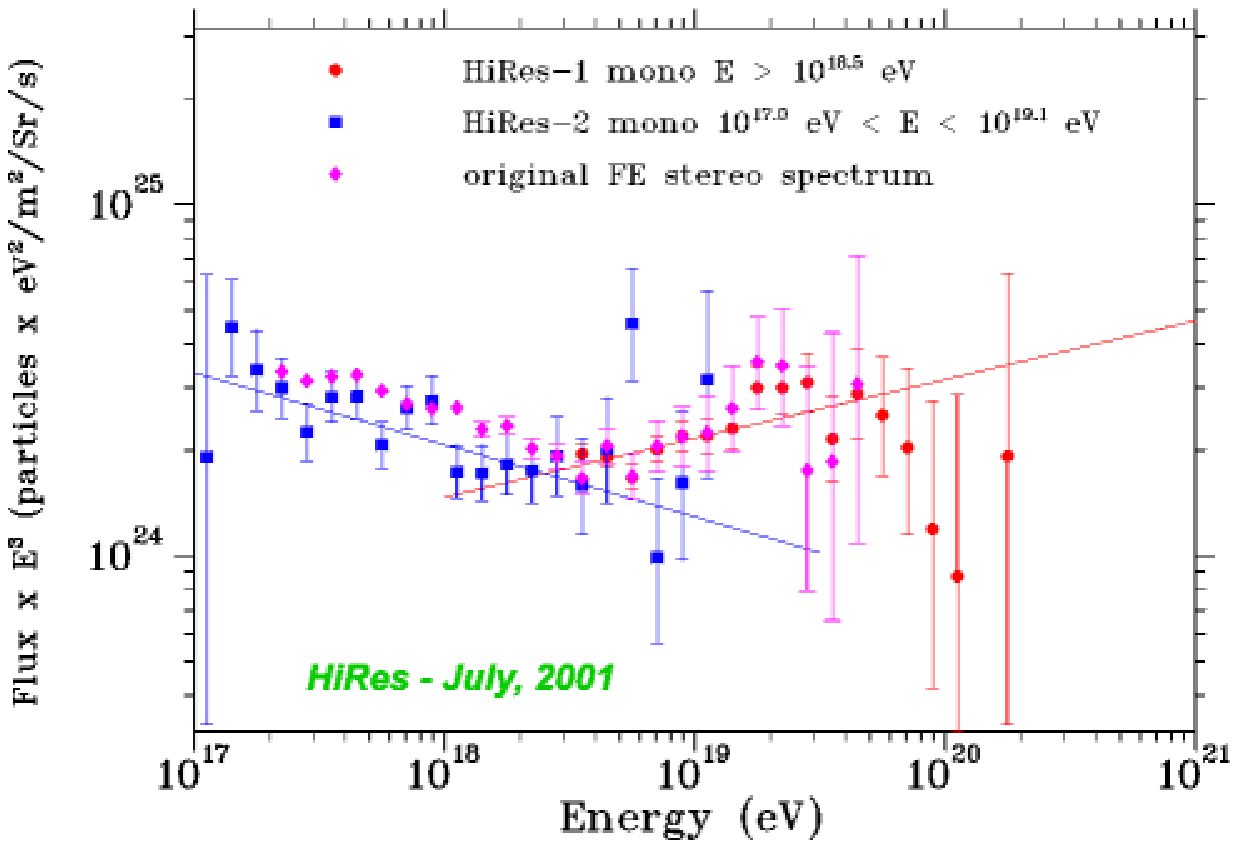}
\caption{As in Fig.\,2, but as measured by the HiRes experiment.}
\label{three}
\end{figure}

The accuracy of the energy resolution of these experiments is a critical
issue. With a particle flux of order 1 event per km$^2$ per
century, these events are studied by using the earth's
atmosphere as a particle detector. The experimental signature of an
extremely high-energy cosmic particle is
a shower initiated by the particle. The primary particle creates an
electromagnetic and
hadronic cascade.  The electromagnetic shower grows to a shower maximum,
and is subsequently absorbed by the
atmosphere.

The shower can be observed by: i)
sampling the electromagnetic and hadronic components when they reach
the ground
with an array of particle detectors such as scintillators, ii)
detecting the fluorescent
light emitted by atmospheric nitrogen excited by the passage of the
shower particles, iii) detecting the Cerenkov light emitted by the
large number
of particles at shower maximum, and iv)~detecting muons and neutrinos
underground.

The bottom line on energy measurement is that, at this time, several
experiments using the first two techniques agree on the energy of
EeV-showers within a typical resolution of 25\%. Additionally, there is a
systematic error of order 10\% associated with the modeling of the
showers. All techniques are indeed subject to the ambiguity of particle
simulations that involve physics beyond the LHC. If the final outcome
turns out to be an erroneous inference of the energy of the shower
because
of new physics associated with particle interactions at the
$\Lambda_{\rm{QCD}}$ scale, we will be happy to contemplate this
discovery
instead.

\looseness=-1
Could the error in the energy measurement be significantly
larger than 25\%? The answer is almost certainly
negative. A variety of techniques have been developed to overcome the
fact that conventional air shower arrays do calorimetry by sampling
at a single depth. They also give results within the range already
mentioned. So do the fluorescence experiments that embody continuous
sampling calorimetry. The latter are subject to understanding the
transmission of fluorescent light in the dark night atmosphere --- a
challenging problem given its variation with weather. Stereo
fluorescence detectors will eventually eliminate this last hurdle by
doing
two
redundant
measurements of the same shower from different locations. The HiRes
collaborators have one year of data on tape which should allow them to
settle energy calibration once and for all.

The premier experiments, HiRes and AGASA, agree that cosmic rays with
energy in excess of 10\,EeV are not galactic in origin and that
their spectrum extends beyond 100\,EeV. They disagree on almost
everything else. The AGASA experiment claims evidence that the highest
energy cosmic rays come
from point sources, and
that they are mostly heavy nuclei. The HiRes data do not support
this. Because of such low statistics, interpreting the measured fluxes as
a
function of energy is like reading tea leaves; one cannot help however
reading different messages in the spectra (see Fig.\,2 and Fig.\,3).

\subsection{The Highest Energy Cosmic Rays: Fancy}

\subsubsection{Acceleration to $>100$ EeV?}

It is sensible to assume that, in order to accelerate a proton to
energy $E$ in a magnetic field $B$, the size $R$ of the accelerator
must be larger than the gyroradius of the particle:
\begin{equation}
R > R_{\rm gyro} = {E\over B}\,.
\end{equation}
That is, the accelerating magnetic field must contain the particle
orbit. This condition yields a maximum energy
\begin{equation}
E = \gamma BR
\end{equation}
by dimensional analysis and nothing more. The $\gamma$-factor has
been included to allow for the possibility that we may not be at rest
in the frame of the cosmic accelerator.  The result would be the
observation of boosted particle energies.
Theorists' imagination regarding the accelerators has been limited to
dense regions where exceptional gravitational forces
create relativistic particle flows: the dense cores of exploding
stars, inflows on supermassive black holes at the centers of active
galaxies, annihilating black holes or neutron stars. All
speculations involve collapsed objects and we can therefore replace $R$
by the Schwartzschild radius
\begin{equation}
R \sim GM/c^2
\end{equation}
to obtain
\begin{equation}
E \propto \gamma BM \,.
\end{equation}
Given the microgauss magnetic field of our galaxy, no structures are
large or massive enough to reach the energies of the highest energy
cosmic rays. Dimensional analysis therefore limits their sources to
extragalactic objects; a few common speculations are listed in
Table\,1.

Nearby active galactic nuclei, distant by $\sim100$~Mpc and
powered by a billion solar mass black holes, are candidates.
With kilogauss fields, we reach 100\,EeV. The jets (blazars) emitted by
the
central black hole could reach similar energies in accelerating
substructures (blobs) boosted in our direction by Lorentz factors of 10
or
possibly higher. The neutron star or black hole remnant of a
collapsing
supermassive star could support magnetic fields of $10^{12}$\,Gauss,
possibly larger. Highly relativistic shocks with $\gamma > 10^2$
emanating
from the
collapsed black hole could be the origin of gamma ray bursts and,
possibly, the source of the highest energy cosmic rays.

\begin{table*}[t]
\caption{Requirements to generate the highest energy cosmic rays in
astrophysical sources.}
\centering\leavevmode
\begin{tabular}{|llll|}
\hline
\multicolumn{4}{|c|}{Conditions with $E \sim 10\rm\ EeV$}\\
\hline
$\bullet$\ Quasars& $\gamma\cong 1$& $B\cong 10^3$ G& $M\cong 10^9
M_{\rm sun}$\\
$\bullet$\ Blazars& $\gamma\gsim 10$& $B\cong 10^3$ G& $M\cong 10^9
M_{\rm sun}$\\
$\bullet$\ \parbox[t]{1.0in}{Neutron Stars\\ Black Holes\\ $\vdots$}&
$\gamma\cong 1$& $B\cong 10^{12}$ G& $M\cong M_{\rm sun}$\\
$\bullet$\ GRB& $\gamma\gsim 10^2$& $B\cong 10^{12}$ G& $M\cong M_{\rm
sun}$\\
\hline
\end{tabular}
\end{table*}

The above speculations are reinforced by the fact that the sources
listed are also the sources of the highest energy gamma rays
observed. At this point, however, a
reality check is in order. The above
dimensional analysis applies to the Fermilab accelerator: 10 kilogauss
fields over several kilometers corresponds to 1\,TeV. The argument holds
because, with optimized design and perfect alignment of magnets, the
accelerator
reaches efficiencies matching the dimensional limit. It is highly
questionable that nature can achieve this feat. Theorists can
imagine acceleration in shocks with an efficiency of perhaps 10\%.

The astrophysics problem of obtaining such high-energy particles is so
daunting that many believe that
cosmic rays are not the beams of cosmic accelerators but the decay
products of remnants from the early Universe, such as topological
defects associated with a Grand Unified Theory (GUT) phase transition.

\subsubsection{Are Cosmic Rays Really Protons: the GZK Cutoff?}

All experimental signatures agree on the particle nature of the
cosmic rays --- they look like protons or, possibly, nuclei. We
mentioned at the beginning of this article that the Universe is
opaque to photons with energy in excess of tens of TeV because they
annihilate into electron pairs in interactions with the cosmic microwave 
background.
Protons also interact with background
light, predominantly by photoproduction of the $\Delta$-resonance,
i.e.\ $p + \gamma_{CMB} \rightarrow \Delta \rightarrow \pi + p$ above
a threshold energy $E_p$ of about 50\,EeV given by:
\begin{equation}
2E_p\epsilon > \left(m_\Delta^2 - m_p^2\right) \,.
\label{eq:threshold}
\end{equation}
The major source of proton energy loss is photoproduction of pions on
a target of cosmic microwave photons of energy $\epsilon$. The
Universe is, therefore, also opaque to the highest energy cosmic rays,
with an absorption length of
\begin{eqnarray}
\lambda_{\gamma p} &=& (n_{\rm CMB} \, \sigma_{p+\gamma_{\rm
CMB}})^{-1}\\
&\cong& 10\rm Mpc,
\end{eqnarray}
when their energy exceeds 50\,EeV. This
so-called GZK cutoff establishes a universal upper limit on
the energy of the cosmic rays. The cutoff is robust,
depending only on two known numbers: $n_{\mathrm{CMB}} = 400\rm\,cm^{-3}$
and
$\sigma_{p+\gamma_{\mathrm{CMB}}} = 10^{-28}\rm\,cm^2$
  \cite{gzk1,gzk2,gzk3,gzk4}.

Protons with energy in excess of 100\,EeV, emitted in distant quasars
and gamma ray bursts, will lose their energy to pions before
reaching our detectors. They have, nevertheless, been observed, as we
have
previously discussed. They do not point to any sources within the
GZK-horizon however, i.e. to sources in our local cluster of
galaxies. There are three possible resolutions: i) the protons are
accelerated in nearby sources, ii)~they do reach us from distant sources
which accelerate them to even higher energies than we observe, thus
exacerbating the acceleration problem, or iii) the highest energy cosmic
rays are not protons.

The first possibility raises the challenge of finding an appropriate
accelerator by confining these already unimaginable sources to our
local galactic cluster. It is not impossible that all cosmic rays are
produced by the active galaxy M87, or by a nearby gamma ray burst
which exploded a few hundred years ago.

Stecker \cite{stecker2} has speculated that the highest energy cosmic
rays are Fe nuclei with a delayed GZK cutoff. The details are
complicated but the relevant quantity in the problem is $\gamma=E/AM$,
where A is the atomic number and M the nucleon mass. For a fixed
observed energy, the smallest boost above GZK threshold is associated
with the largest atomic mass, i.e.~Fe.

\begin{figure}[t!]
\centering\leavevmode
\includegraphics[width=5in]{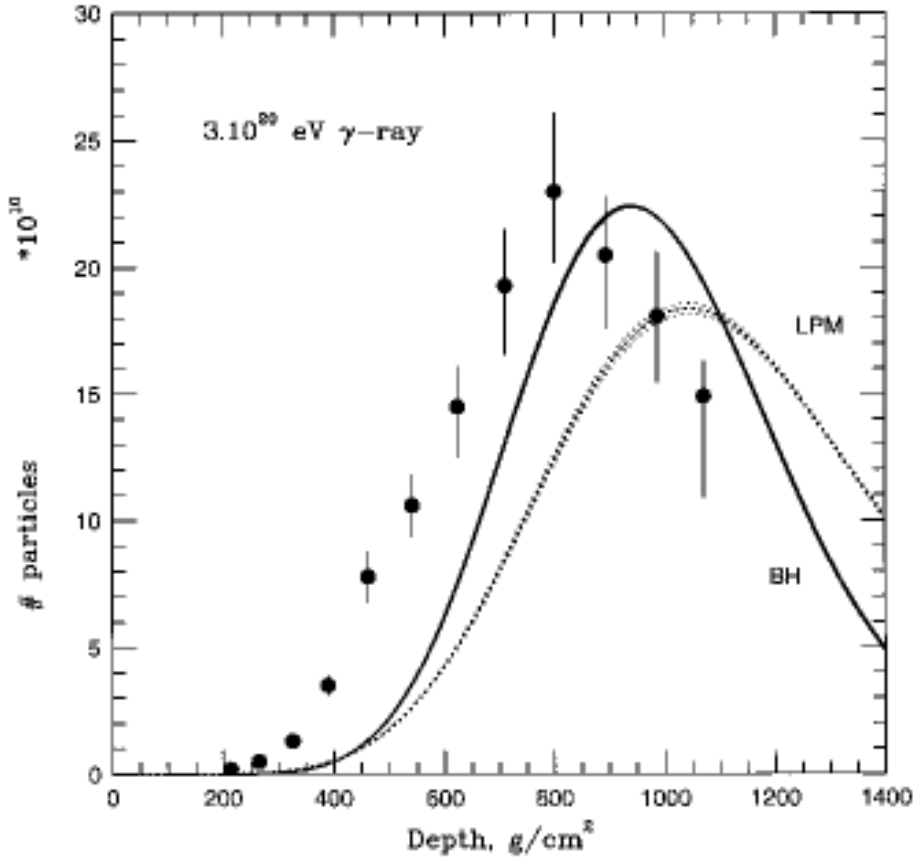}
\caption{The composite atmospheric shower profile of a $3\times
10^{20}$\,eV gamma ray shower calculated with Landau-Pomeranchuk-Migdal 
(dashed) and
Bethe-Heitler (solid)
electromagnetic cross sections. The central line shows the average shower
profile and the upper and lower lines show 1~$\sigma$ deviations --- not
visible for the BH case, where lines overlap. The experimental shower
profile
is shown with the data points. It does not fit the profile of a photon
shower.}
\label{four}
\end{figure}

\subsubsection{Could Cosmic Rays be Photons or Neutrinos?}

Topological defects predict that the highest energy cosmic rays are
predominantly photons. A topological defect will suffer a chain decay
into GUT particles X and Y, that subsequently
decay to familiar weak bosons, leptons and quark or gluon jets. Cosmic
rays are, therefore, predominately the fragmentation products of these
jets. We know from accelerator studies that, among the fragmentation
products of jets, neutral pions (decaying into photons) dominate, in
number, protons by close to two orders of magnitude. Therefore, if the
decay of topological defects is the source of the highest energy cosmic
rays, they must be photons. This is a problem because there is
compelling evidence that the highest energy cosmic rays are not photons:

1. The highest energy event observed by Fly's Eye is not likely to
be a photon \cite{vazquez}.  A photon of 300\,EeV will interact with the
magnetic field of the earth far above the atmosphere and disintegrate
into lower energy cascades --- roughly ten at this particular energy.
The detector subsequently collects light produced by the fluorescence of
atmospheric nitrogen along the path of the high-energy showers
traversing the atmosphere. The anticipated shower profile of a 300\,EeV
photon
is shown in Fig.\,4. It disagrees with the data. The observed shower
profile does fit that of a primary proton, or,
possibly, that of a nucleus. The shower profile information is
sufficient,
however, to conclude that the
event is unlikely to be of photon origin.

2. The same conclusion is
reached for the Yakutsk event that is characterized by a huge number
of secondary muons, inconsistent with an electromagnetic cascade
initiated by a gamma ray.

3. The AGASA collaboration claims evidence
for ``point" sources above 10\,EeV. The arrival directions are
however smeared out in a way consistent with primaries deflected by
the galactic magnetic field. Again, this indicates charged primaries
and excludes photons.

4. Finally, a recent reanalysis of the Haverah Park disfavors photon
origin of the primaries \cite{WatsonZas}.

Neutrino primaries are definitely ruled out. Standard model neutrino
physics is understood, even for EeV energy. The average $x$ of
the parton mediating the neutrino interaction is of
order $x \sim \sqrt{M_W^2/s} \sim 10^{-6}$ so that the perturbative
result for the neutrino-nucleus cross section is calculable from measured
HERA structure functions. Even at 100\,EeV a
reliable value of the cross section can be obtained based on QCD-inspired
extrapolations of the structure function. The neutrino cross section is
known to better than an order of magnitude. It falls 5 orders of
magnitude
short of the strong cross sections required to make a neutrino interact
in
the upper atmosphere to create an air shower.

Could EeV neutrinos be strongly interacting because of new physics?
In theories with TeV-scale gravity, one can imagine that graviton
exchange dominates all interactions and thus erases the difference
between
quarks and
neutrinos at the energies under consideration. The actual models
performing this feat require a fast turn-on of the cross
section with energy that violates S-wave unitarity 
\cite{han1,han2,han3,han4,han5,han6,han7,han8,han9,ring1,ring2}.

We have exhausted the possibilities.  Neutrons, muons and other
candidate primaries one may think of are unstable. EeV neutrons
barely live long enough to reach us from sources at the edge of our
galaxy.

\subsection{A Three Prong Assault on the Cosmic Ray Puzzle}

We conclude that, where the highest energy cosmic rays are concerned,
both
the
accelerator mechanism and the particle physics are enigmatic. The mystery
has inspired a worldwide effort to tackle the
problem with novel experimentation in three complementary areas of
research: air shower detection, atmospheric Cerenkov astronomy and
underground neutrino astronomy. While some of the future instruments
have additional missions, all are likely to have a major impact on cosmic
ray physics.

\subsubsection{Giant Cosmic Ray Detectors}

With super-GZK fluxes of the order of a single event per square 
kilometer, per century, the outstanding problem is the lack of
statistics; see Fig.\,2 and Fig.\,3. In the next five years, a
qualitative
improvement can be expected from the operation of the HiRes fluorescence
detector in Utah.
With improved instrumentation yielding high quality
data from 2 detectors operated in coincidence, the interplay between
sky transparency and energy
measurement can be studied in detail. We can safely anticipate that
the existence of super-GZK cosmic rays will be conclusively
demonstrated by using the instrument's calorimetric
measurements. A mostly Japanese collaboration has proposed a
next-generation fluorescence detector, the Telescope Array.

The Auger air shower array is confronting the low rate problem with a
huge
collection area covering 3000 square kilometers on an elevated plain
in Western Argentina. The instrumentation consists of 1600 water
Cerenkov detectors spaced by 1.5\,km. For calibration, about 15
percent of the showers occurring at night will be viewed by 3
HiRes-style fluorescence detectors. The detector is expected to observe
several
thousand events per year above 10\,EeV and tens above 100\,EeV. Exact
numbers will depend on the detailed shape of the observed
spectrum which is, at present, a matter of speculation.

\subsubsection{Gamma rays from Cosmic Accelerators}

\begin{figure}[t!]
\centering\leavevmode
\includegraphics[width=5in]{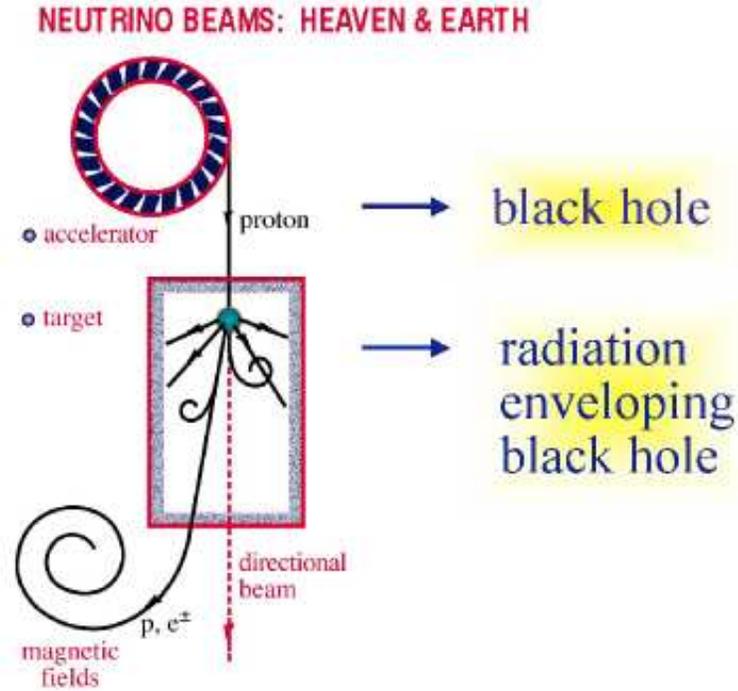}
\caption{Diagram of cosmic accelerator and beam dump.  See text for
discussion.}
\label{five}
\end{figure}


An alternative way to identify the source(s) of the highest energy 
cosmic rays
is
illustrated in Fig.\,5. The cartoon draws our attention to the fact that
cosmic accelerators are also cosmic beam dumps which produce secondary
photon and neutrino beams.
Accelerating particles to TeV energy and above requires relativistic,
massive bulk flows. These are likely to originate from the
exceptional gravitational forces associated with dense
cores of exploding stars, inflows onto supermassive black holes
at the centers of active galaxies, annihilating black holes or
neutron stars. In such situations, accelerated particles
are likely to pass through intense radiation fields or dense
clouds of gas surrounding the black hole. This leads to the production of
secondary
photons and neutrinos that accompany the primary cosmic ray beam.
An example of an electromagnetic beam dump is the UV radiation field that
surrounds the central black hole of active galaxies.
The target material, whether a gas of particles or of photons,
is likely to be tenuous enough that the primary beam
and the photon beam are only partially attenuated. However, shrouded
sources from which only neutrinos can emerge, as in
terrestrial beam dumps at CERN and Fermilab, are also a possibility.

The astronomy event of the 21st century could be the simultaneous
observation of TeV-gamma rays, neutrinos and gravitational waves from
cataclysmic events associated with the source of the highest energy 
cosmic rays.

We first concentrate on the possibility of detecting high-energy
photon beams. After two decades, ground-based gamma ray astronomy has
become a mature science 
\cite{weekes,weekes2,weekes3,weekes4,weekes5,weekes6}.  A large mirror,
viewed by an
array of photomultipliers, collects the Cerenkov light emitted by air
showers and images the showers in order to determine the arrival
direction and the nature of the primary particle.
These
experiments have opened a new window in astronomy by extending the
photon spectrum to 20\,TeV, and possibly beyond. Observations have
revealed spectacular TeV-emission from galactic supernova remnants
and nearby quasars, some of which emit most of their energy in very
short bursts of TeV-photons.

But there is the dog that didn't bark. No evidence has emerged
for the $\pi^0$ origin of TeV radiation. Therefore, no cosmic
ray sources have yet been identified. Dedicated searches for photon beams
from
suspected cosmic ray sources, such as the supernova remnants IC433
and $\gamma$-Cygni, came up empty handed. While not relevant to the
topic covered by this paper, supernova remnants are theorized to be the
sources of the bulk of the cosmic rays that are of galactic origin.
However, the evidence is still circumstantial.

The field of gamma ray astronomy is buzzing with activity to
construct second-generation instruments. Space-based detectors are
extending their reach from GeV to TeV energy with AMS and,
especially, GLAST, while the ground-based Cerenkov collaborations are
designing instruments with lower thresholds. Soon, both techniques should
generate overlapping measurements in the $10-10^2$~GeV energy range. All
ground-based air Cerenkov experiments aim
at lower threshold, better angular and energy resolution, and a
longer duty cycle. One can, however, identify three pathways to reach
these goals:

\begin{enumerate}

\item
larger mirror area, exploiting the parasitic use of solar collectors
during nighttime (CELESTE, STACEY and SOLAR\,II) \cite{pare},
\item
better, or rather, ultimate imaging with the 17m MAGIC
mirror, \cite{magic}
\item
larger field of view and better pointing and energy measurement using
multiple telescopes (VERITAS, HEGRA and HESS).

\end{enumerate}

The Whipple telescope pioneered the atmospheric Cerenkov technique.
VERITAS \cite{veritas} is an array of 9 upgraded Whipple telescopes,
each with a
field of view of 6 degrees. These can be operated in coincidence for
improved angular resolution, or be pointed at 9 different 6 degree
bins in the night sky, thus achieving a large field of view. The
HEGRA collaboration \cite{hegra} is already operating four telescopes
in coincidence and is building an upgraded facility with excellent
viewing and optimal location near the equator in Namibia.

There is a dark horse in this race: Milagro \cite{milagro}.  The
Milagro idea is to lower the threshold of conventional air shower arrays
to 100~GeV by
instrumenting a pond of five million gallons of ultra-pure water with
photomultipliers. For time-varying signals, such as bursts, the
threshold may be even lower.

\subsubsection{Neutrinos from Cosmic Accelerators}

How many neutrinos are produced in association with the cosmic ray beam?
The answer to this question, among many 
others \cite{snowmass1,snowmass2},  provides the rational for building 
kilometer-scale neutrino detectors.

Let's first consider the question for the accelerator beam producing 
neutrino
beams at an accelerator laboratory. Here the target absorbs all parent
protons as well as the muons, electrons and gamma rays (from $\pi^0
\rightarrow \gamma + \gamma$)
produced. A pure neutrino beam exits the dump. If nature constructed 
such a ``hidden source" in the heavens,
conventional astronomy will not reveal it. It cannot be the source of the
cosmic rays, however, for which the dump must be partially transparent to
protons.

In the other extreme, the accelerated proton interacts, thus
producing the observed high-energy gamma rays, and subsequently escapes 
the dump. We refer to this as a transparent source.  Particle physics 
directly relates the number of
neutrinos to the number of observed cosmic rays and gamma 
rays\cite{halzenzas}. Every observed cosmic
ray
interacts
once, and only once, to produce a neutrino beam determined only by
particle physics. The neutrino flux for such a transparent cosmic ray
source is referred to as the
Waxman-Bahcall flux \cite{wb1,wb2,R1,R2} and is shown as the horizontal
lines labeled ``W\&B" in
Fig.\,6. The calculations is valid for $E\simeq100$\,PeV. If the flux is calculated at both lower and higher cosmic ray energies, however, larger values are found. This is shown as the non-flat line labeled ``transparent" in Fig.\,6. On
the lower side, the neutrino flux is higher because it is normalized to a
larger cosmic ray flux. On the higher side, there are more cosmic
rays in the dump to produce neutrinos because the observed flux at Earth
has been
reduced by absorption on microwave photons, the GZK-effect. The increased
values of the neutrino flux are also shown in Fig.\,6. The gamma ray 
flux of $\pi^0$ origin associated with a transparent source is 
qualitatively at the level of observed flux of non-thermal TeV gamma 
rays from individual sources\cite{halzenzas}.

Nothing prevents us, however, from imagining heavenly beam dumps with target densities somewhere between those of hidden and transparent sources. When
increasing the target photon density, the proton beam is absorbed in the
dump and the number of neutrino-producing protons is enhanced relative to
those escaping the source as cosmic rays. For the extreme source of this
type, the observed cosmic rays are all decay products of neutrons
with larger mean-free paths in the dump. The flux for such a source is
shown as the upper horizontal line in Fig.\,6.

\begin{figure}[t!]
\centering\leavevmode
\includegraphics[width=5in]{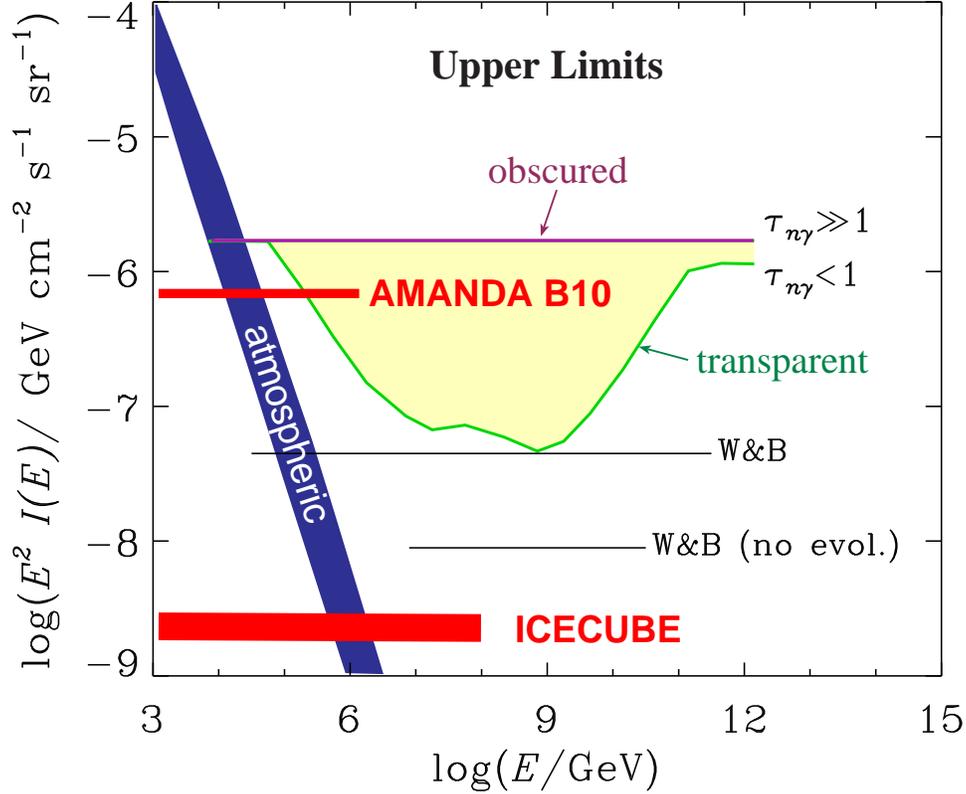}
\caption{The neutrino flux from compact astrophysical accelerators.
Shown
is the range of possible neutrino fluxes associated with the the highest
energy cosmic rays.  The lower line, labeled ``transparent", represents a source where each cosmic
ray interacts only once before escaping the object.  The upper line, labeled ``obscured", represents an ideal neutrino source where all cosmic rays escape in the
form of neutrons.  Also shown is the ability of AMANDA and IceCube to
test
these models.}
\label{six}
\end{figure}

The above limits are derived from the fact that theorized neutrino
sources
do not overproduce cosmic rays. Similarly, observed gamma ray fluxes
constrain potential neutrino sources because for every parent charged
pion
($\pi^{\pm}\rightarrow l^{\pm}+\nu$), a neutral pion and two gamma rays
($\pi^0\rightarrow \gamma + \gamma$) are produced. The
electromagnetic
energy associated with the
decay of neutral pions should not exceed observed astronomical fluxes.
These calculations must take into account cascading of the
electromagnetic
flux in the background photon and magnetic fields. A simple
argument relating high-energy photons and neutrinos produced by secondary
pions can still be derived by relating their total energy and allowing
for
a steeper photon flux as a result of cascading. Identifying the photon 
fluxes with those of non-thermal TeV
photons emitted by supernova remnants and blazers, we predict neutrino 
fluxes at the same level as the Waxman-Bahcall flux. It is important to 
realize however that there is no evidence
that these are the decay products of $\pi^0$'s. The sources
of the cosmic rays have not been revealed by photon or proton astronomy 
\cite{gas1,gas2,gas3,gas4}.

For neutrino detectors to succeed they must be sensitive to the range of
fluxes covered in Fig.\,6. The AMANDA detector has already entered the
region of sensitivity and is eliminating specific models which predict 
the largest neutrino fluxes within the range of values allowed by general
arguments. The IceCube detector, now under construction, is sensitive to 
the full range of beam dump models,
whether generic as or modeled as active galaxies or gamma ray bursts. 
IceCube will reveal the
sources of the cosmic rays or derive an upper limit that will
qualitatively raise the bar for solving the cosmic ray puzzle. The
situation could be nothing but desperate with the escape to top-down
models being cut off by the accumulating evidence that the highest energy
cosmic rays are not photons. In top-down models, decay products 
predominantly
materialize as quarks and gluons that materialize as jets of neutrinos
and
photons and very few protons. We will return to top-down models at the 
end of this review.

\section{High-energy Neutrino Telescopes}

\begin{figure}[t!]
\centering\leavevmode
\includegraphics[width=5in]{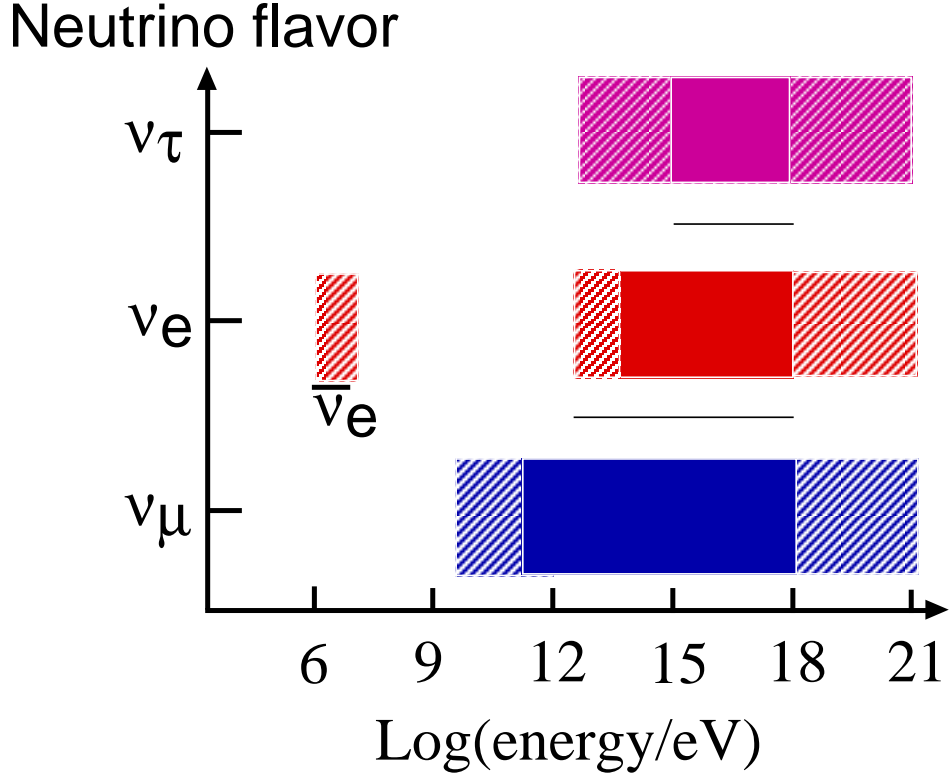}
\caption{Although IceCube detects neutrinos of any flavor, at TeV-EeV 
energies, it can identify their flavor and measure
their energy in the ranges shown. Filled areas: particle identification,
energy, and angle. Shaded areas: energy and angle.}
\label{seven}
\end{figure}

\subsection{Observing High-energy Neutrinos}

Although details vary from experiment to experiment, high-energy neutrino
telescopes consist of strings of photo-multiplier tubes (PMT)
distributed
throughout a natural Cerenkov medium such as water or ice.  Typical
spacing of PMT is 10-20 meters along a string with string spacing of
30-100 meters.  Such experiments can observe neutrinos of different 
flavors over a wide
range of energies by using a variety of methods:
\begin{itemize}

\item{} Muon neutrinos that interact via charged current interactions
produce a muon (along with a visible hadronic shower if the neutrino is
of
sufficient energy).  The muon travels through the medium
producing Cerenkov radiation which is detected by an array of PMT. The
timing, amplitude (number of Cerenkov photons) and topology of the PMT
signals is used to reconstruct the muon's
path. The muon energy threshold for such a reconstruction is typically in
the range of 10-100 GeV.

To be detected, a neutrino must interact via charged current and produce
a muon with sufficient range to reach the detector.  The probability of
detection is therefore the product of the interaction probability (or
the inverse interaction length $\lambda_{\nu}^{-1}=n\sigma_{\nu}$) and 
the
range of the muon $R_{\mu}$:
\begin{eqnarray}
P_{\nu \rightarrow \mu} \simeq  n \sigma_\nu R_\mu
\end{eqnarray}
where $n$ is the number density of target nucleons, $\sigma_\nu$ is the
charged current interaction cross section \cite{crossgandhi} and the range is $R_\mu\simeq 
5\,\rm{m}\,\rm{per}\, \rm{GeV}\,$for low energy muons. The muon range is determined 
by catastrophic energy
loss (brehmsstrahlung, pair production and deep inelastic scattering)
for muons with energies exceeding $\sim 500\,$GeV \cite{snr1,learned}.

\item{} Muon, tau or electron neutrinos which interact via charged or
neutral current interactions produce showers which can be observed when
the interaction occurs within or close to the detector volume. Even the 
highest
energy showers penetrate water or ice less than 10\,m, a distance
short compared to the typical spacing of the PMT. The Cerenkov light
emitted by shower particles, therefore, represents a point source of 
light as viewed by the array. The radius over which PMT signals are produced is
250\,m for a 1\,PeV shower; this radius grows or decreases by
approximately 50\,m with every decade of shower energy.  The threshold
for
showers is generally higher than for muons which
limits neutral current identification for lower energy neutrinos.  The
probability for a neutrino to interact within the detector's effective
area and to generate a
shower within its volume is approximately given by:
\begin{eqnarray}
P_{\nu \rightarrow \mu} \simeq  n \sigma_\nu L
\end{eqnarray}
where $\sigma_\nu$ is the charged+neutral current interaction cross
section, $L$ is the length of the detector along the path of the neutrino
and $n$, again, is the number density of target nucleons.

\item{} Tau neutrinos are more difficult to detect but produce 
spectacular signatures at PeV energies.  The identification of charged
current tau neutrino events is made by observing one of two
signatures:  double bang events \cite{double1,double2,double3} and
lollypop
events \cite{pdd, lollypop}.  Double bang events occur when a tau lepton is 
produced
along with a hadronic shower in a charged current interaction within the
detector volume and the tau decays producing a electromagnetic or
hadronic shower before exiting the detector (as shown in Fig.\,8).  Below
a few PeV, the two showers cannot be distinguished.  Lollypop events
occur
when only the second of the two showers of a double bang event occurs 
within the detector volume and a tau lepton track is identified entering
the shower over several hundred meters. The incoming $\tau$ can be 
clearly distinguished from a muon. A muon initiating a PeV shower would 
undergo observable catastrophic energylosses. Lollypop events are useful 
only at several PeV energies are
above.  Below this energy, tau tracks are not long enough to be
identified.

A feature unique to tau neutrinos is that they are not depleted in
number by absorption in the earth.  Tau neutrinos which interact 
producing a tau
lepton generate another tau neutrino when the tau lepton decays, thus
only degrading the energy of the neutrino \cite{taudep1,taudep2, taudep3, taudep4}.

\item{} Although MeV scale neutrinos are far below the energies required
to identify individual events, large fluxes of MeV electron 
anti-neutrinos
interacting via charged current could be detected by observing higher
counting rates of individual PMT over a time window of several
seconds.  The enhancement rate in a single PMT will be buried in dark
noise of that PMT.  However, summing the signals from all PMT over a 
short time window can reveal significant excesses, for instance form a 
galactic supernova.

\end{itemize}
With these signatures, neutrino astronomy can study neutrinos from the
MeV
range to the highest known energies ($\sim10^{20}$eV).

\begin{figure}[t!]
\centering\leavevmode
\includegraphics[width=5in]{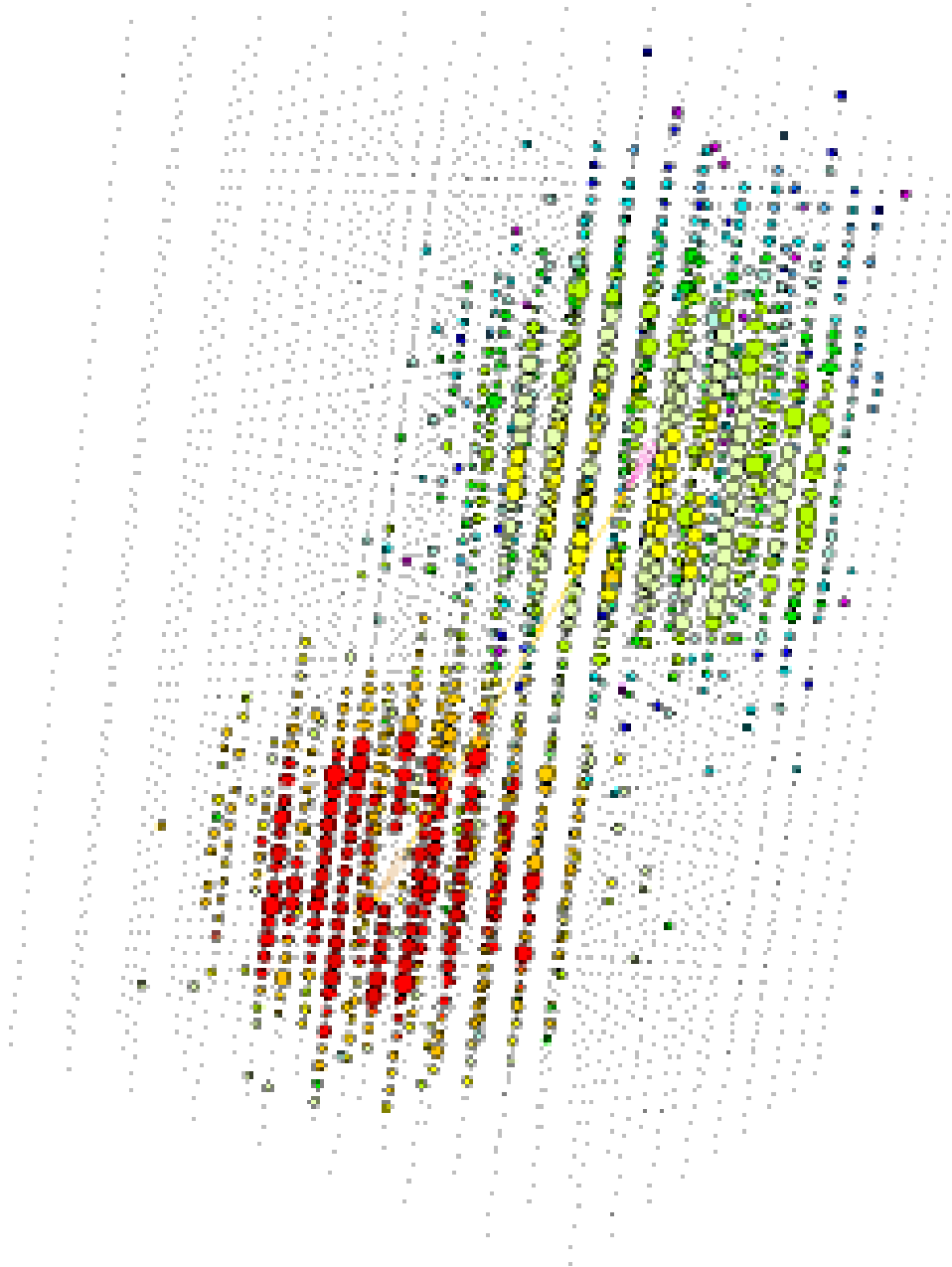}
\caption{Simulation of an ultra high-energy tau lepton generated by the
interaction of a 10 PeV tau neutrino (first shower), followed by the
decay
of the secondary tau lepton (second shower). The shading represents the
time
sequence of the hits. The size of the dots
corresponds to the number of photons detected by the individual
photomultipliers.}
\label{sixteen}
\end{figure}

\subsection{Large Natural Cerenkov Detectors}

A new window in astronomy is upon us as high-energy neutrino telescopes
see first light \cite{light}.  Although neutrino telescopes have multiple
interdisciplinary science
missions, the search for the sources of the highest-energy cosmic
rays stands out because it most directly identifies the size of the
detector required to do the science \cite{snr1,learned}.  For
guidance in estimating
expected signals, one makes use of data covering the highest-energy
cosmic rays in Fig.\,2 and Fig.\,3 as well as known sources of
non-thermal,
high-energy gamma
rays. Estimates based on this information suggest that a kilometer-scale 
detector is needed to see neutrino
signals as previously discussed.

The same conclusion is reached using specific models. Assume, for
instance, that gamma ray bursts (GRB) are the cosmic
accelerators of the highest-energy cosmic rays. One can calculate
from textbook particle physics how many neutrinos are produced when
the particle beam coexists with the observed MeV energy photons in
the original fireball. We thus predict the observation of 10--100
neutrinos of PeV energy per year in a detector with a square kilometer 
effective area.  GRB are an example of a generic beam
dump associated with the highest energy cosmic rays.  We will work
through this example in some detail in later sections. In general, the 
potential scientific
payoff of
doing neutrino astronomy arises from the great penetrating power of
neutrinos, which allows them to emerge from dense inner regions of
energetic sources.

The strong scientific motivations for a large area, high-energy neutrino
observatory lead to the formidable challenges of developing effective,
reliable and affordable detector technology.  Suggestions to use a large
volume of deep ocean water for high-energy neutrino
astronomy were made as early as the 1960s. Today, with the first 
observation of neutrinos in
the Lake Baikal and
the South Pole neutrino telescopes, there is optimism
that the technological challenges of building neutrino telescopes have
been met.

Launched by the bold
decision
of
the DUMAND collaboration to construct such an instrument, the first 
generation of neutrino telescopes is
designed to reach a large telescope area and detection volume for a
neutrino
threshold of order 10~GeV \cite{dumand,dumand1,dumand2}. This relatively
low threshold permits calibration of
the novel instrumentation on the known flux of atmospheric neutrinos.
The
architecture is optimized for reconstructing the Cerenkov light
front radiated
by an up-going, neutrino-induced muon.  Up-going
muons must be identified in a background of down-going, cosmic ray muons
which
are more than $10^5$ times more frequent for a depth of $\sim$1--2
kilometers. The earth is used
as a filter to screen out the background of down-going cosmic ray
muons. This makes
neutrino detection possible over the hemisphere of sky faced by the
bottom
of the detector.

\begin{figure}[t!]
\centering\leavevmode
\includegraphics[width=5in]{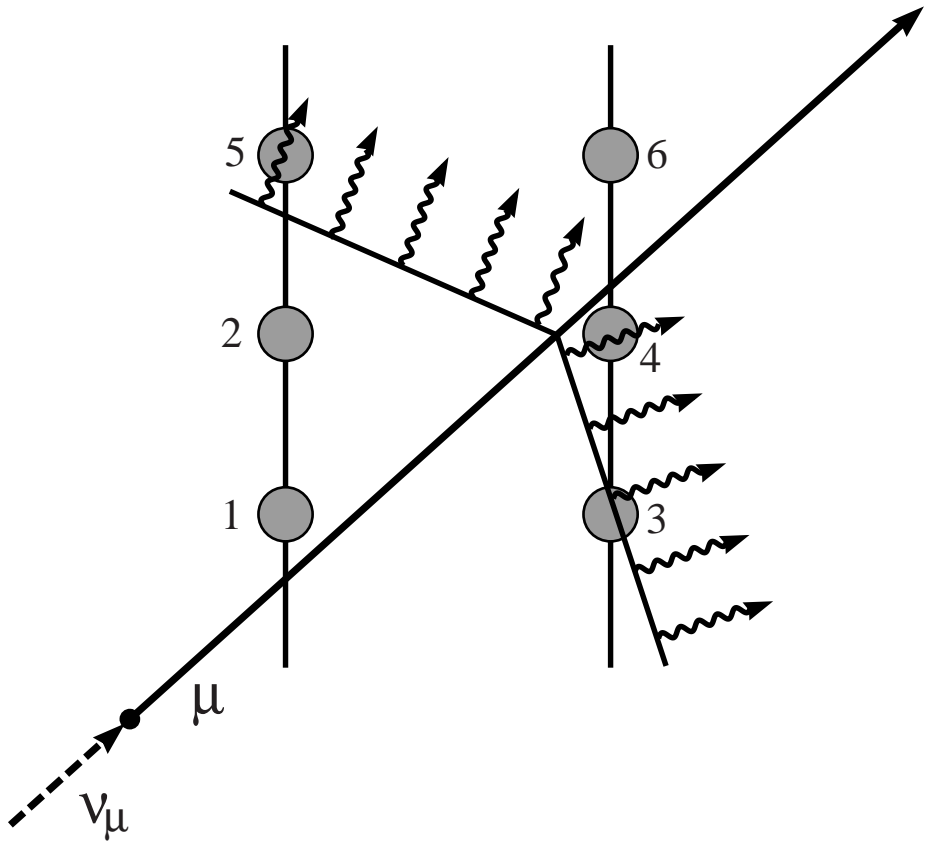}
\caption{The arrival times of the Cerenkov photons in 6 optical sensors
determine the direction of the muon track.}
\label{eight}
\end{figure}

The optical requirements on the detector medium are severe. A large
absorption
length is needed because it determines the required spacing of the
optical
sensors
and, to a significant extent, the cost of the detector. A long scattering
length is needed to preserve the geometry of the Cerenkov
pattern. Nature has
been kind and offered ice and water as natural Cerenkov
media. Their
optical properties are, in fact, complementary. Water and ice have
similar
attenuation length, with the roles of scattering and absorption
reversed. Optics seems, at present, to drive the evolution of ice and
water
detectors in predictable directions: towards very large telescope area in
ice
exploiting the long absorption length, and towards lower threshold and
good
muon track reconstruction in water exploiting the long scattering length.

\subsubsection{Baikal, ANTARES, Nestor and NEMO:  Northern Water}

Whereas the science is compelling, we now turn to the challenge of 
developing effective detector technology. With the termination of
the
pioneering DUMAND experiment, the efforts in water are, at present,
spearheaded
by the Baikal experiment \cite{baikal,baikal1,baikal2,baikal3}. The
Baikal
Neutrino Telescope is
deployed in Lake Baikal, Siberia, 3.6\,km from shore at a depth of
1.1\,km. An
umbrella-like frame holds 8 strings, each instrumented with 24 pairs of
37-cm
diameter {\it QUASAR} photomultiplier tubes. Two PMT are
required to trigger in coincidence in order to suppress the large 
background rates produced by natural
radioactivity and bioluminescence in individual PMT. Operating with 144 
optical modules
(OM) since
April 1997, the {\it NT-200} detector was completed in April 1998 with
192 OM. Due to unstable electronics only $\sim 60$
channels took data during 1998. Nevertheless 35
neutrino-induced up-going muons were identified in the first 234 live
days of data; see Fig.\,10 for a 70 day sample. The neutrino events are
isolated from the cosmic ray muon background
by imposing a restriction on the chi-square of the fit of measured 
photon arrival times and amplitudes to a Cherenkov cone, and by
requiring consistency between the reconstructed trajectory and the
spatial
locations of the OMs reporting signals. In order to guarantee a minimum
lever
arm for track fitting, they only consider events with a projection of
the most distant channels on the track larger than 35 meters. This does,
of
course, result in a higher energy threshold. Agreement with the expected
atmospheric neutrino flux of 31 events shows that the Baikal detector
is understood. Stability and performance of the detector have
improved in 1999 and 2000 data taking \cite{baikal3}.

The Baikal site is competitive with deep oceans, although the smaller
absorption length of Cerenkov light in lake water requires a
somewhat denser spacing of the OMs. This does, however, result in a
lower threshold which is a definite
advantage, for instance for oscillation measurements and WIMP searches. 
They
have shown that their shallow depth of 1 kilometer does not represent a
serious
drawback. A significant advantage is that the site has a seasonal ice
cover which allows reliable and inexpensive deployment and repair of
detector
elements.

\begin{figure}[t!]
\centering\leavevmode
\includegraphics[width=5in]{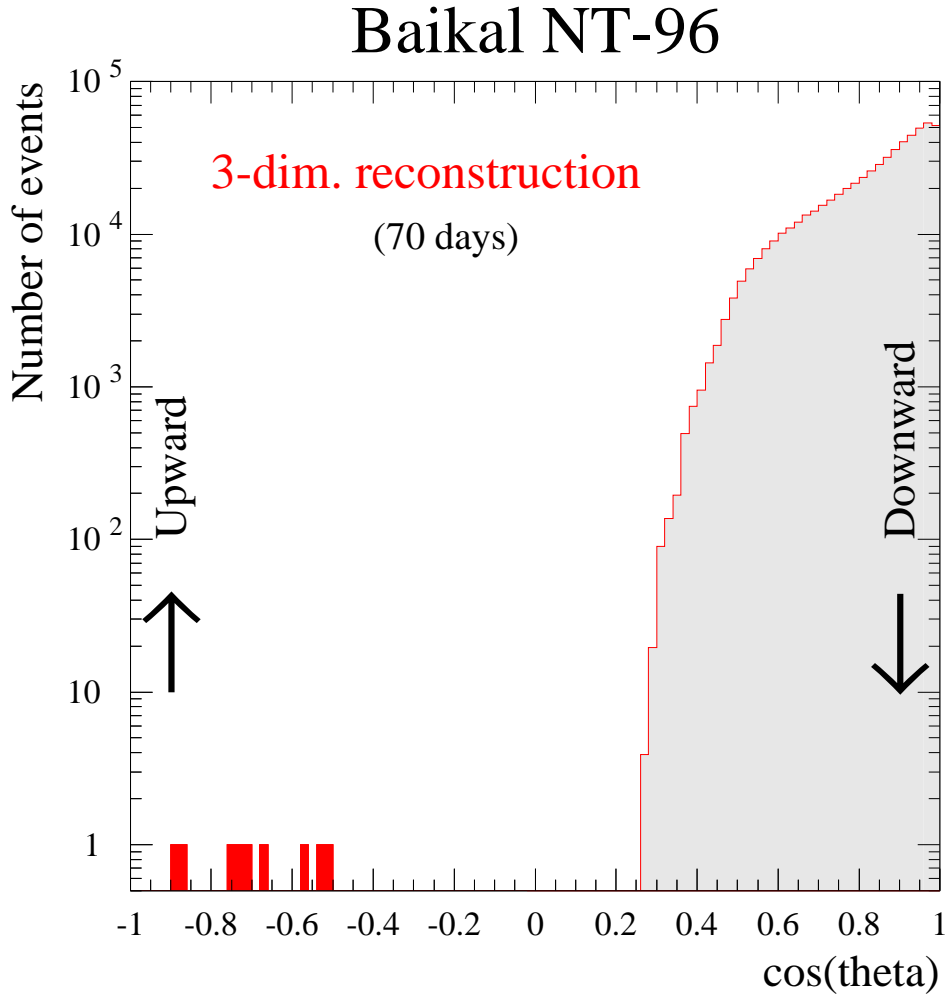}
\caption{Angular distribution of muon tracks in the Lake Baikal
experiment
after the cuts described in the text.}
\label{nine}
\end{figure}

In the following years, {\it NT-200} will be operated as a neutrino
telescope
with an effective area between $10^3$ and $5\times 10^3$\,m$^2$,
depending
on
energy. Presumably too small to detect neutrinos from extraterrestrial
sources,
{\it NT-200} will serve as the prototype for a larger telescope. For
instance,
with 2000 OMs, a threshold of  $10$ to $20$\,GeV and an effective area of
$5\times10^4$ to $10^5$\,m$^2$, an expanded Baikal telescope could fill
the gap
between present underground detectors and planned high threshold
detectors
of
cubic kilometer size. Its key advantage would be low energy threshold.

The Baikal experiment represents a proof of concept for future deep ocean
projects that have the advantage of larger depth and optically superior
water. Their
challenge is to find reliable and affordable solutions to a variety of
technological challenges for deploying a deep underwater detector.
Several
groups are confronting the problem; both NESTOR and ANTARES are
developing
rather different detector concepts in the Mediterranean.

The NESTOR collaboration \cite{nestor,nestor1,nestor2}, as part of a
series
of ongoing
technology tests, is testing the umbrella structure which will hold the
OMs.
They have already deployed two aluminum ``floors", 34\,m in diameter,
to a
depth of 2600\,m. Mechanical robustness was demonstrated by towing the
structure, submerged below 2000\,m, from shore to the site and back.
These
tests
should soon be repeated with two fully instrumented floors. The cable 
connecting the instrument to the counting house on shore has been 
deployed. The final detector
will consist of a tower of 12 six-legged floors vertically separated by
30\,m.
Each floor contains 14 OMs with four times the photocathode area of the
commercial 8~inch photomultipliers used by AMANDA and ANTARES.

The detector concept is patterned along the Baikal design. The symmetric
up/down orientation of the OMs will result in uniform angular acceptance
and
the relatively close spacings will result in a low energy
threshold. NESTOR does have the
advantage of a superb site off the coast of Southern Greece, possibly
the best in the Mediterranean. The
detector can be deployed below 3.5\,km relatively close to shore. With
the
attenuation length peaking at 55\,m near 470\,nm, the site is optically
similar
to that of the best deep water sites investigated for neutrino astronomy.

The ANTARES collaboration \cite{antares,antares1,antares2} is currently
constructing a
neutrino telescope at a 2400\,m deep Mediterranean site off Toulon,
France. The site is a trade-off between acceptable optical properties
of the water and easy access to ocean technology. Their detector
concept requires remotely operated vehicles for making
underwater connections. Results on water quality are very encouraging
with an absorption length of 40\,m at 467\,nm and 20\,m at 375\,nm,
and a scattering
length exceeding 100\,m at both wavelengths. Random noise, exceeding
50\,khz per OM, is eliminated by
requiring coincidences between neighboring OMs, as is done in the Lake
Baikal
design. Unlike other water experiments, they will point all
photomultipliers
sideways or down in order to avoid the effects of biofouling. The problem
is
significant at the Toulon site, but only affects the upper pole region of
the
OM. Relatively weak intensity and long duration bioluminescence results
in
an
acceptable deadtime of the detector. They have demonstrated their
capability to
deploy and retrieve a string, and have reconstructed down-going muons 
with
8~OMs deployed on the test string.

The ANTARES detector will consist of 13 strings, each equipped with
30 stories and 3~PMT per story. This detector will have an area of
about $3\times 10^4\rm\,m^2$ for 1~TeV muons --- similar to AMANDA-II
--- and is planned to be fully deployed by the end of 2004. The
electro-optical cable linking the underwater site to the shore was
successfully deployed in October 2001.

NEMO, a new R\&D initiative based in Catania, Sicily has been mapping
Mediterranean
sites, studying mechanical structures and low power electronics. One
hopes
that with a successful pioneering neutrino detector of $10^{-3}\rm\,
km^3$
in
Lake Baikal and a forthcoming $10^{-2}\rm\, km^3$ detector near Toulon,
the
Mediterranean effort will converge on a $10^{-1}\rm\, km^3$ detector,
possibly at the
NESTOR site \cite{spiro,NEMO}. For neutrino astronomy to become a viable
science,
several projects will have to succeed in addition to AMANDA.
Astronomy, whether in the optical or in any other wave-band, thrives on a
diversity of complementary instruments, not on ``a single best
instrument".

\subsubsection{AMANDA: Southern Ice}

Construction of the first-generation AMANDA-B10
detector \cite{B4,nature00,amanda2,amanda3,amanda4} was completed in the
austral summer 96--97. It consists of 302 optical modules deployed
at a depth
of 1500--2000~m; see Fig.\,11. Here the optical modules consist of 8-inch
photomultiplier tubes and are controlled by passive electronics. Each is 
connected to the surface by a
cable that transmits the high voltage as well as the anode current of a 
triggered photomultiplier. The instrumented volume and the effective 
telescope area
of
this instrument matches those of the ultimate DUMAND Octagon detector
which,
unfortunately, could not be completed.


Depending on depth, the absorption length of blue  and UV
light in the ice varies between 85 and 225 meters.  The
effective scattering length, which
      combines the mean-free  path
$\lambda$ with the average scattering angle $\theta$ as
$\lambda   \over (1-\langle cos\theta \rangle )$,
varies from 15 to 40 meters \cite{science}.
      Because the absorption length
of light in the ice is very long and the scattering length
relatively short, many photons are delayed by scattering. In
order to reconstruct the muon track, maximum
likelihood methods are used, which take into  account the scattering
and absorption of photons as determined from calibration
measurements  \cite{B4}. A Bayesian formulation of the
likelihood \cite{ghill1}, which accounts for the much larger rate of
down-going cosmic-ray muon tracks relative to up-going signal, has been
particularly effective in decreasing the chance for a
down-going muon to be misreconstructed as up-going.

\begin{figure}[t!]
\centering\leavevmode
\includegraphics[width=5in]{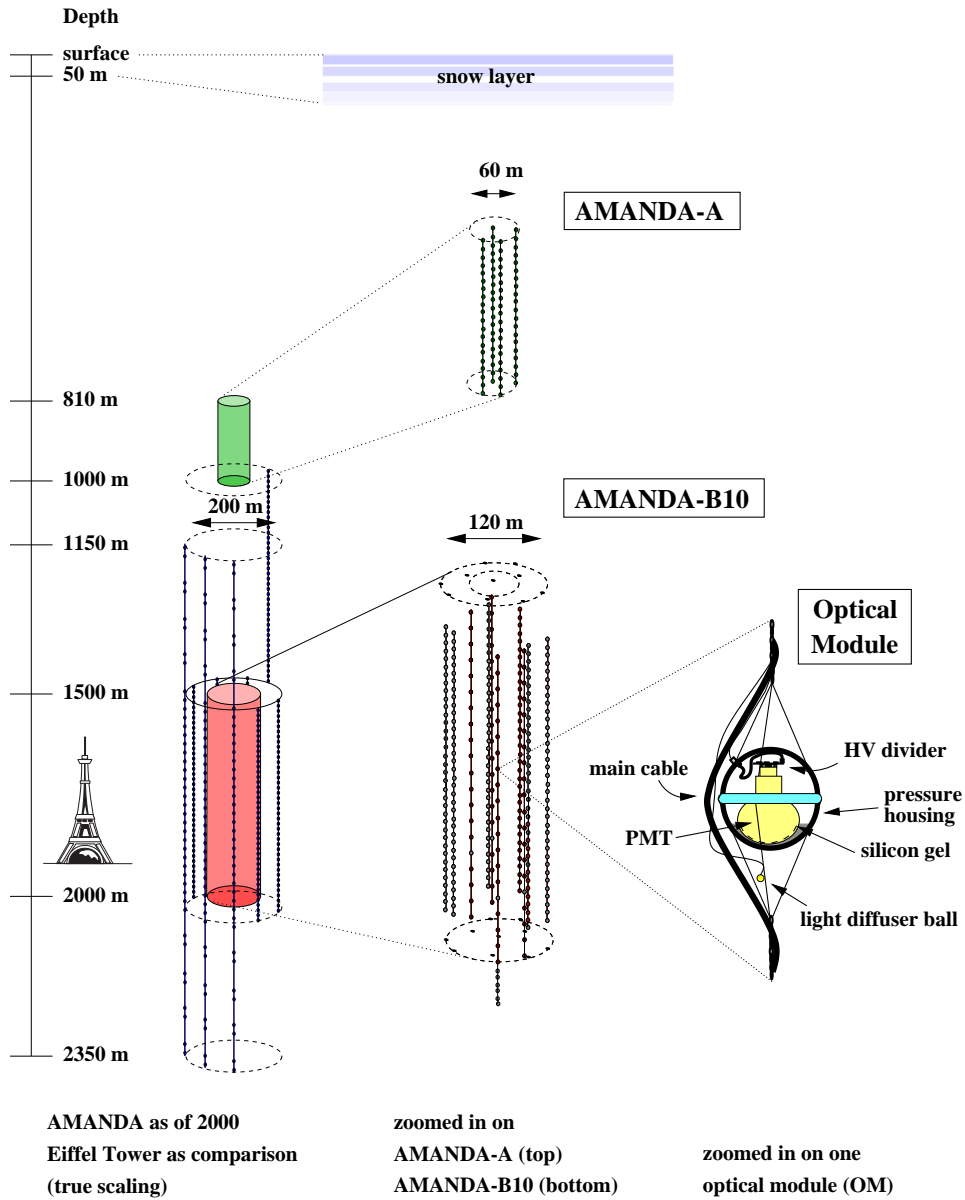}
\caption{The AMANDA detector and a schematic diagram of
an optical module.   Each dot represents an optical module.
The modules are separated by 20 meters in the inner strings 1-4,
and by 10 meters in the outer strings 5-10.}
\label{ten}
\end{figure}

Other types of events that might appear to be up-going muons
must also be considered and eliminated.  Rare cases, such as
muons which undergo catastrophic energy loss, for instance
through bremsstrahlung, or that  are coincident with other muons,
must be investigated.
To
this end, a series of requirements or quality criteria,
based on the characteristic time and spatial pattern of
photons associated with a muon track and the response of the
detector, are applied to all events that, in the first
analysis, appear to be up-going muons.
      For example, an event which has a large number of optical modules
hit by photons unscattered (relative to the expected Cerenkov times of
the
reconstructed track) has a
high quality.
       By making these requirements (or ``cuts'')
increasingly selective, they eliminate more of
the background of false up-going events while still
retaining a significant fraction of the true up-going muons,
i.e., the neutrino signal.  Two different and independent
analyses of the same data covering 138 days of
observation in 1997 have been undertaken. These analyses yielded
comparable
numbers of up-going muons (153 in analysis A, 188 in analysis B).
Comparison of these results with their respective Monte Carlo simulations
shows that they are consistent with each other in terms of the numbers of
events, the number of events in common, and, as discussed below, the
expected properties of atmospheric neutrinos.

In Fig.~12, from analysis A, the
experimental events are compared to simulations of
background and signal as a function of the (identical)
quality requirements placed on the three types of events:
experimental data, simulated up-going muons from atmospheric
neutrinos, and a simulated background of down-going
cosmic ray muons.
      For simplicity
in presentation, the levels of the individual types of cuts
have been combined into a single parameter representing the
overall event quality, and the comparison is made in the
form of ratios.
Fig.\,12 shows events for which the quality level is 4 and
higher.  As the quality level is increased further, the
ratios of simulated background to experimental data and
experimental data to simulated signal both continue their
rapid decrease, the former toward zero and the latter toward
unity.  Over the same range, the ratio of experimental data
to the simulated sum of background and signal remains near
unity.  At an event quality of 6.9 there are 153 events in
the sample of experimental data and the ratio to predicted
signal is 0.7.  The conclusions are that (1) the quality
requirements have reduced the events from misreconstructed
down-going muons in the experimental data to a negligible
fraction of the signal and that (2) the experimental data
      behave in the same way as the simulated
atmospheric neutrino signal for events
      that pass the stringent cuts. They estimate that
the remaining signal is contaminated by instrumental background at $15
\pm
7$ percent.

\begin{figure}[t!]
\centering\leavevmode
\includegraphics[width=5in]{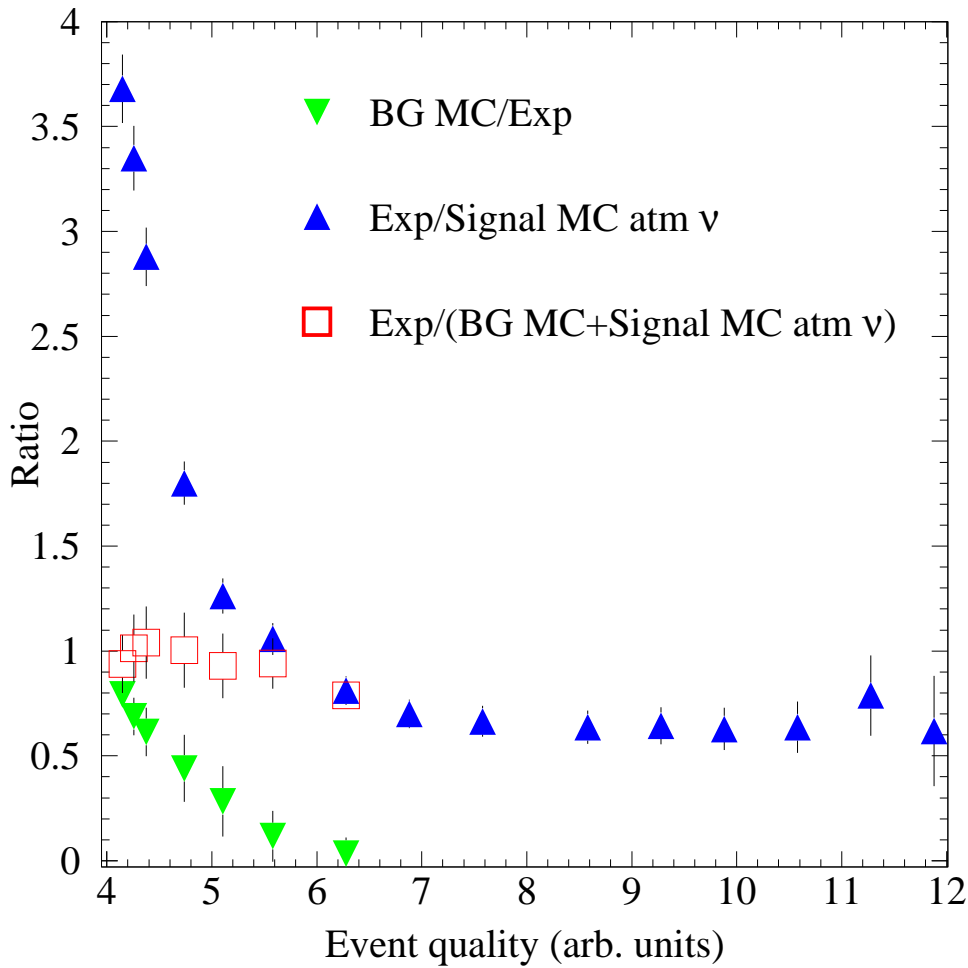}
\caption{Reconstructed muon events in AMANDA-B10 are compared to
simulations of background cosmic ray muons (BG MC) and
simulations of atmospheric neutrinos (Signal MC atm $\nu$) as a
function of ``event quality'', a variable indicating the
severity of the cuts designed to enhance the signal. Note
that the comparison is made in the form of ratios.}
\label{eleven}
\end{figure}

The estimated uncertainty on the number of events predicted
by the signal Monte Carlo simulation (which includes
uncertainties in the high-energy atmospheric neutrino flux,
the sensitivity of the optical modules, and the
precise optical properties of the ice) is +40\% to $-$50\%.  The
observed ratio of experiment to simulation (0.7) and the
expectation (1.0) therefore agree within errors.

The shape of the zenith angle distribution from analysis B is
compared to a simulation of the atmospheric neutrino signal
in Fig.~13 in which the two distributions have been
normalized to each other.  The variation of the measured
rate with zenith angle is reproduced by simulation to
within the statistical uncertainty.  Note that the tall geometry
of the detector strongly influences the dependence on zenith
angle in favor of more vertical muons.

\begin{figure}[t!]
\centering\leavevmode
\includegraphics[width=5in]{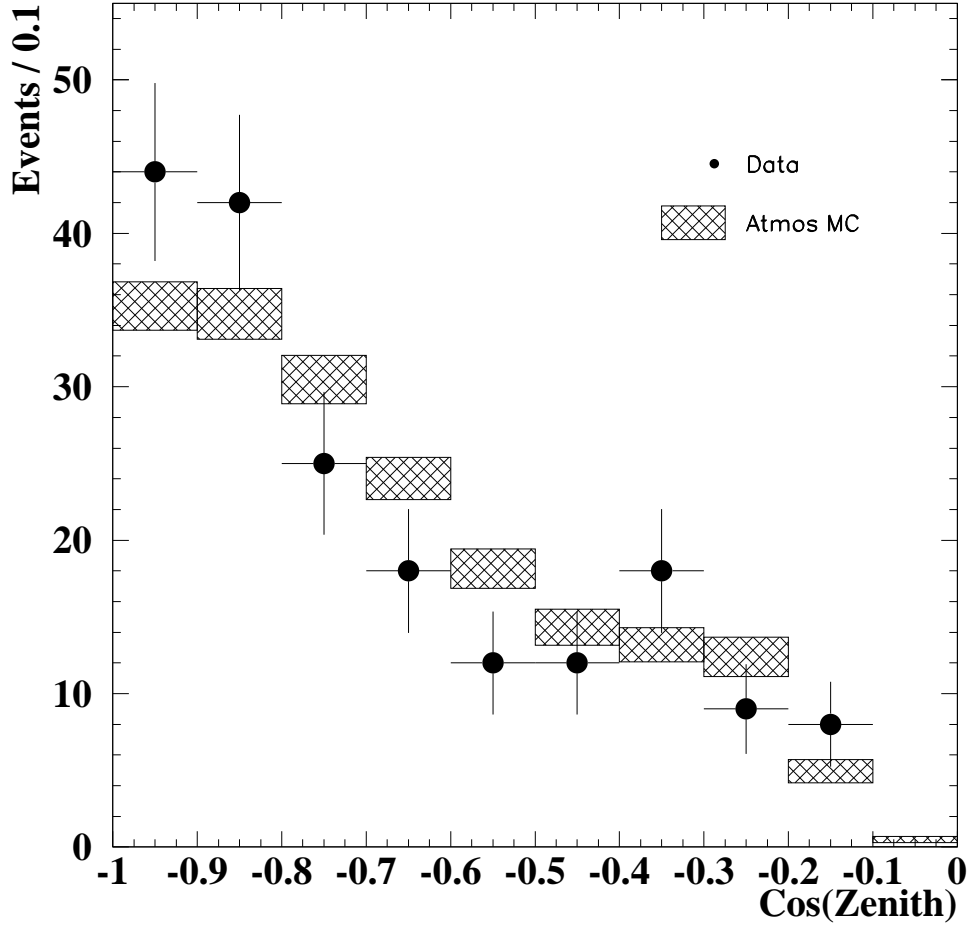}
\caption{Reconstructed zenith angle distribution for AMANDA-B10.  The
points mark the data and the shaded boxes a simulation of
atmospheric neutrino events. The widths of the boxes
indicate the error bars. The overall normalization of the
simulation has been adjusted to match the data.}
\label{twelve}
\end{figure}

Estimates of the energies of the up-going muons (based on
simulations of the number of optical modules that
participate in an event) indicate that the energies of these
muons are  in the range from 100 GeV to $\sim 1~\rm{TeV}$.  This is
consistent with their atmospheric neutrino origin.

The agreement between simulation and experiment shown in
Fig.\,12 and 13, taken together with other comparisons of
measured and simulated events, leads us to conclude that the up-going
muon events observed by AMANDA are produced mainly by
atmospheric neutrinos.

The arrival directions of the neutrinos observed in both
analyses are shown in Fig.~14.
       A statistical analysis
indicates no evidence for point sources in this sample.  An
estimate of the energies of the up-going muons indicates that all events
have energies consistent with an
      atmospheric neutrino origin.  This corresponds to a level
of sensitivity to a diffuse flux of high-energy extra-terrestrial
neutrinos of order $
dN/dE_{\nu} = 10^{-6} E_{\nu}^{-2} \rm\, cm^{-2}\, s^{-1}\,
sr^{-1}\,   GeV^{-1}, $ assuming an  $E^{-2}$
spectrum \cite{ghill2}. This upper limit excludes a
variety of theoretical models which assume the hadronic
origin of TeV photons from active galaxies and blazars.
Searches for neutrinos from gamma ray bursts, magnetic
monopoles, and for a cold dark matter signal from the center
of the Earth yield limits comparable to or better
than those from smaller underground neutrino detectors that
have operated for a much longer period.

\begin{figure}[t!]
\centering\leavevmode
\includegraphics[width=5in]{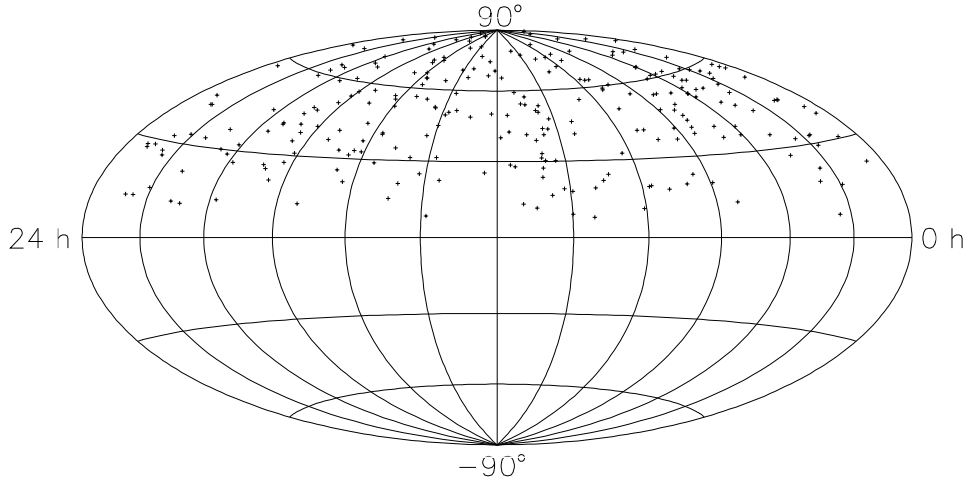}
\caption{Distribution in declination and right ascention of the up-going
AMANDA-B10 events on the sky.}
\label{thirteen}
\end{figure}

Data are being taken now with the larger
array, AMANDA-II consisting of an additional 480 OMs.

\subsubsection{IceCube: A Kilometer-Scale Neutrino Observatory}

The IceCube project \cite{icecube1,icecube2} at the South Pole is a
logical
extension of the
research and
development work performed over the past several years by the AMANDA
Collaboration. The optimized design for IceCube is an array of
4800
photomultiplier tubes each enclosed in a transparent pressure
sphere to comprise an optical module similar to those in AMANDA.
In the IceCube design, 80~strings are regularly spaced by 125~m over
an area of approximately one square kilometer, with OMs at depths from
1.4 to 2.4~km below the surface.  Each string consists of OMs
connected electrically and mechanically to a long cable which brings
OM signals to the surface.  The array is deployed one string at a
time. For each string, a enhanced hot-water drill melts a hole in the 
ice to
a depth of about 2.4~km in less than 2 days. The drill is then removed 
from the hole and a
string with 60~OMs vertically spaced by 17~m is deployed before the water
re-freezes.  The signal cables from all the strings are brought to a
central location which houses the data acquisition electronics, other
electronics, and computing equipment.

Each OM contains a 10~inch PMT that detects individual photons of
Cerenkov
light generated in the optically clear ice by muons and electrons
moving with velocities near the speed of light.

Background events are mainly down-going muons from
cosmic ray interactions in the atmosphere above the detector. The
background is monitored for
calibration purposes and background rejection by the IceTop air
shower array covering the detector.

Signals from the optical modules are digitized and transmitted to the
surface such that a photon's time of arrival at an OM can be determined
to within less than 5 nanoseconds. The electronics at the surface 
determines
when an event has occurred (e.g., that a muon traversed or passed near
the array) and records the information for subsequent event
reconstruction and analysis.

At the South Pole site
(see Fig.\,15),
a computer
system accepts the data from the event trigger via the data
acquisition system. The event rate, which is dominated by down-going
cosmic ray muons, is estimated to be 1--2~kHz. The technology that
will be employed in IceCube has been developed,
tested, and demonstrated in AMANDA deployments, in laboratory testing,
and in simulations validated by AMANDA data. This includes the
instrument architecture, technology, deployment, calibration, and
scientific utilization of the
proposed detector.  There have been yearly improvements in the AMANDA
system, especially in the OMs, and in the overall quality of the
information obtained from the detector.  In the 1999/2000 season, a
string was deployed with optical modules containing readout
electronics inside the OM. The information is sent
digitally to
the surface over twisted-pair electrical cable. This option eliminates
the need for optical fiber cables and simplifies calibration of the
detector elements. This digital technology is the baseline technology
of IceCube. For more details, see Ref.\,\cite{web2}.

\begin{figure}[t!]
\centering\leavevmode
\includegraphics[width=5in]{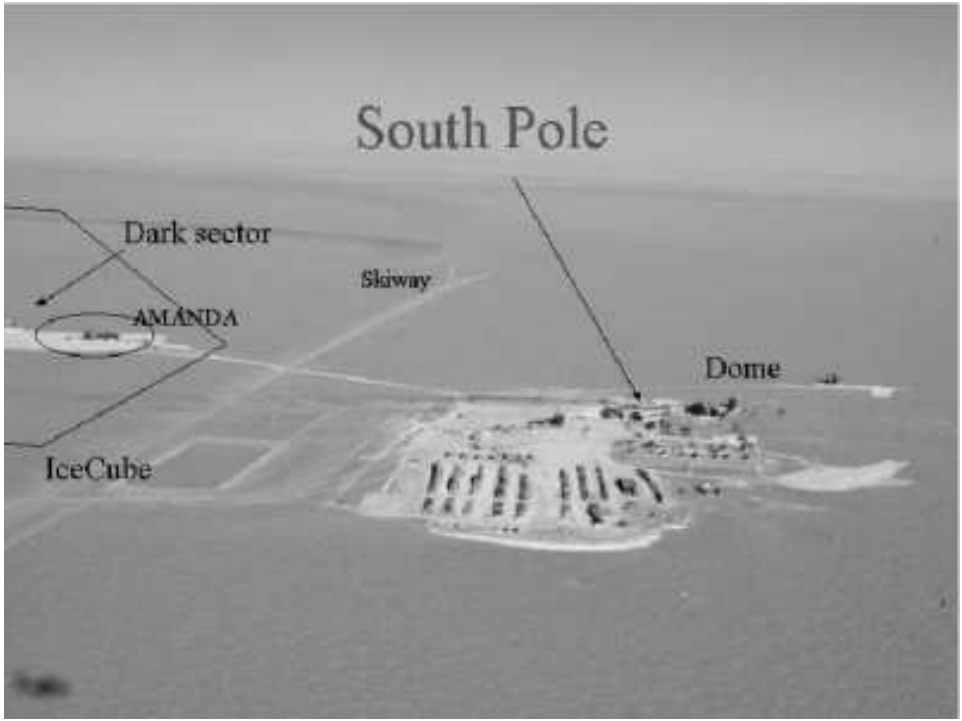}
\caption{The South Pole site, showing the residential dome
and associated buildings,
             the skiway where planes land, the dark sector with the Martin
A. Pomerantz Observatory in which
             the AMANDA electronics are housed, and a rough outline of
where
IceCube strings are to be placed.}
\label{fourteen}
\end{figure}

The construction of neutrino telescopes is overwhelmingly motivated by
their discovery potential in astronomy, astrophysics, cosmology and
particle physics.  To maximize this potential, one must design an
instrument with the largest possible effective telescope area to
overcome the neutrino's small cross section with matter, and the best
possible angular and energy resolution to address the wide diversity
of possible signals.

At this point in time, several of the new instruments (such as the
partially deployed Auger array, HiRes, Magic, Milagro and AMANDA~II) are
less than one year from delivering results. With rapidly growing
observational capabilities, one can realistically hope, almost 100 years 
after their discovery, the puzzling origin of the cosmic rays will be 
deciphered. The solution will almost certainly reveal
unexpected astrophysics or particle physics.

\subsection{EeV Neutrino Astronomy}

At extremely high energies, new techniques can be used to
detect astrophysical neutrinos.  These include the detection of acoustic
and radio signals induced by super-EeV neutrinos interacting in water,
ice or salt domes, or the detection of horizontal air showers by large
conventional cosmic ray experiments such as the Auger array.

Horizontal air showers are likely to be initiated by a neutrino because
showers induced by primary cosmic rays are unlikely to penetrate the
$\sim 36,000\, \rm{g}/\rm{cm}^2$ of atmosphere along the horizon.
Isolated penetrating muons may survive but they can be experimentally
separated from a shower initiated by a neutrino close to the
detector. Horizontal air shower experiments can also use nearby
mountains as a target, e.g. to observe the decay of tau leptons
produced in charged current interactions in the moutain. The sensitivity 
of an air
shower array to detect an ultra high-energy neutrino is described by its
acceptance, expressed in units of $\rm{km}^3$ water equivalent
steradians ($\rm{km}^3 \rm{we\,sr}$).  Typically only showers with zenith
angle greater than $\sim 70\,$ degrees can be identified as neutrinos.
This corresponds to a slant depth of $\sim 2000\, \rm{g}/\rm{cm}^2$.

The acceptance of present air shower experiments, such as AGASA, is
$\sim 1\rm{km}^3 \rm{we\,sr}$ above $10^{10}\,$GeV, and significantly 
less
at lower energies. Auger will achieve ten times greater acceptance at
$10^9\,$\,GeV and 50 times greater near
$10^{12}\,$\,GeV. Nitrogen fluorescence experiments also have the
capability to detect neutrinos as nearly horizontal air showers with
space-based experiments such as EUSO and OWL extending the reach of
Auger. At this point we should point out however that the actual event
rates of these experiments are similar to those for IceCube.
Although IceCube's energy resolution saturates at EeV energies, the
neutrinos are still detected with rates competitive with the most
ambitious horizontal air shower experiments; for a more
detailed comparison see Ref.\,\cite{eevicecube,eevicecube2} .

Radio Cerenkov experiments detect the Giga-Hertz pulse radiated by
shower electrons produced in the interaction of neutrinos in ice. Also,
the
moon, viewed by ground-based radio
telescopes, has been used as a target \cite{moon}. Above a threshold of 
$\simeq1$\,PeV, the large
number of low energy($\simeq MeV$) photons in a shower will produce an 
excess of
electrons over positrons by removing electrons from atoms by Compton
scattering. These are the sources of coherent radiation at radio
frequencies, i.e. above $\sim100$\,MHz. The mechanism is now well
understood. The characteristics and the power of the pulses have been
measured by dumping a photon beam in sand \cite{sand}. The results agree
with calculations \cite{sand2}.

While many proposals exist, the most extensive effort to develop a radio
neutrino detector is RICE (Radio Ice Cerenkov Experiment), which is 
located in the shallow ice above the AMANDA
detector \cite{rice}. It consists of an 18-channel array of radio
receivers distributed within a $8\times 10^6 \, \rm{m}^3 \,$ volume.
The receivers, buried in the ice at depths of 100-300 meters, are
sensitive over the range of 0.2-1 GHz, roughly corresponding to electron
neutrinos with energy of several PeV and above. The ANITA collaboration 
proposes to fly a balloon-borne array of radio antennas on a circular flight over 
Antarctica. ANITA will detect
earth-skimming neutrinos \cite{feng} producing signals emerging from the ice along
the horizon \cite{anita}. With higher threshold but also greater
effective area than RICE (about 1 million $\rm{km}^2$), ANITA should be
sensitive to GZK neutrinos after a lucky 30 day flight (or 3 normal
flights of 10
days).

EeV neutrino-induced showers can also be detected by acoustic emission
resulting from local heating of a dense medium. Existing arrays of
hydrophones, built in the earth's oceans for military application, could
be used for the hydro-acoustic detection of neutrinos with extremely
high energies; for a recent review see \cite{acoustic}.

\section{Cosmic Neutrino Sources}
\subsection{A List of Cosmic Neutrino Sources}

We have previously discussed generic cosmic ray producing beam dumps and
their associated neutrino fluxes.  We now turn to specific sources
of high-energy neutrinos.  The list of proposed sources is long and
includes, but is not limited to:
\begin{itemize}
\item{Gamma Ray Bursts (GRB)} 

GRB, outshining the entire universe for the duration of the burst, are 
perhaps the best
motivated source for high-energy neutrinos \cite{waxmanbahcall, mostlum2,
mostlum3}.  Although we do not yet understand the internal mechanisms
that generate GRB, the relativistic fireball model provides us with a
successful phenomenology accommodating observations.  It is very likely
that GRB are
generated in some type of cataclysmic process involving dying massive 
stars.
GRB may prove to be an excellent source of neutrinos with energies from
MeV to EeV and above. As we shall demonstrate further on, their fluxes
can be calculated in a relatively model independent fashion.

\item{Other Sources Associated with Stellar Objects}

Other theorized neutrino sources associated with compact objects include 
supernova remnants exploding into the
interstellar medium \cite{snr1,snr2,snr3,learned}, X-ray
binaries \cite{snr1,xray1,xray2,xray3}, microquasars 
\cite{learned,micro1,micro2}
and even the sun \cite {snr1,sun1,sun2,learned}, any of which could
provide observable fluxes of high-energy neutrinos.

\item{Active Galactic Nuclei (AGN):  Blazars} 

Blazars, the brightest objects in the universe and the sources of
TeV-energy  gamma rays, have been extensively studied as potential
neutrino sources.  Blazar flares with durations ranging from
months to less than an hour, are believed to be produced by relativistic
jets projected from an extremely massive accreting black hole. Blazars
may be the sources of the highest energy cosmic rays and, in
association, provide observable fluxes of neutrinos from TeV to EeV
energies.

\item{Neutrinos Associated with the Propagation of Cosmic Rays} 

Very high-energy cosmic rays generate neutrinos in interactions
with the cosmic microwave background \cite{cos1,cos2}.  This cosmogenic
flux is among the most likely
sources of high-energy neutrinos, and the most straightforward to
predict.  Furthermore, cosmic rays interact with the Earth's
atmosphere \cite{prompt1, prompt2}
and with the hydrogen concentrated in the galactic plane 
\cite{snr1,plane1,plane2,plane3,learned} producing high-energy
neutrinos.  It has also been proposed that cosmic neutrinos themselves
may produce cosmic rays and neutrinos in interactions with relic
neutrinos $\nu + \nu_b\rightarrow Z$. This is called the Z-burst
mechanism \cite{z1,z2,z3,z4,z5}.

\item{Dark Matter, Primordial Black Holes, Topological Defects and
Top-Down Models} 

The vast majority of matter in the universe is dark with its particle
nature not yet revealed.  The lightest supersymmetric particle, or other 
Weakly Interacting Massive
Particles (WIMPs) propsed as particle candidates for cold dark matter, 
should become gravitationally trapped in the sun, earth or galactic 
center. There, they annihilate generating high-energy neutrinos observable in
neutrino telescopes \cite{ind1,ind2,ind3,ind4,ind5,ind6,ind7}.  Another 
class of dark matter
candidates
are superheavy particles with GUT-scale masses that may generate the 
ultra high-energy cosmic
rays by decay or annihilation, as well as solve the dark matter problem.
These will also generate a substantial
neutrino flux \cite{sh1,sh2,sh3,sh4}.  Extremely high-energy neutrinos
are also predicted in a wide variety of top-down scenarios invoked to
produce cosmic rays, including decaying monopoles, vibrating cosmic
strings \cite{top1,top2} and Hawking radiation
from primordial black holes \cite{bh1,bh2,pbh}.

Any of these sources may or may not provide observable fluxes of 
neutrinos. History testifies to the fact
that we have not been particularly successful at predicting the 
phenomena invariably revealed by new ways of viewing the
heavens. We do, however, know that cosmic rays exist and that nature
accelerates particles to super-EeV energy. In this review we concentrate
on neutrino fluxes associated with the highest energy cosmic rays. Even
here the anticipated flux depends on our speculation regarding the
source. We will work through three much-researched examples: GRB, AGN
and decays of particles or defects associated with the GUT-scale. The
myriad of speculations have been recently reviewed by Learned and
Mannheim \cite{learned}. We concentrate here on neutrino sources 
associated with the observed cosmic rays and gamma rays.

\end{itemize}

\subsection{Gamma Ray Bursts:  A Detailed Example of a Generic Beam Dump}
\subsubsection{GRB Characteristics}

Although there is no such thing as a typical gamma ray burst,
observations of GRB indicate the following common characteristics:
\begin{itemize}

\item GRB are extremely luminous events, often releasing energy of order 
one solar mass in gamma rays.  Typically, $L_{\gamma}\sim 10^{51}$
to $10^{54}$ erg/s is released over durations of seconds or tens of 
seconds.  GRB are
the most luminous sources in the universe.

\item GRB produce a broken power-law spectrum of gamma
rays with $\phi_\gamma\propto E_\gamma^{-2}$ for $E_\gamma \gsim$ 0.1-1
MeV and
$\phi_\gamma\propto E_\gamma^{-1}$ for $E_\gamma \lsim$ 0.1-1 MeV 
\cite{piran,halzenlec}.

\item GRB are cosmological events.  Redshifts exceeding z=4 have been
measured \cite{red1, red2}.

\item GRB are rare.  During it's operation, BATSE observed on average 1
burst per day within its field of view ($\sim1/3$ of the sky).  Assuming
that the
rate of GRB does not significantly change with cosmological time, this
corresponds to one burst per galaxy per million years.  If GRB are
beamed,
they may be more common.

\item GRB produce afterglows of less energetic photons which extend long
after the initial burst \cite{hist4,after2,after3}.

\item The durations of GRB follow a bimodal distribution
with peaks near two seconds and 20 seconds, although some GRB have
durations ranging from milliseconds to 1000 seconds \cite{dur1}.  
Variations in the spectra occur on the scale of milliseconds \cite{msec,dur1} 
is shown in Fig.\,16 \cite{batse}. GRB
afterglows can extend for days \cite{dur1}.

\end{itemize}
\begin{figure}[thb]
\centering\leavevmode
\includegraphics[width=5in]{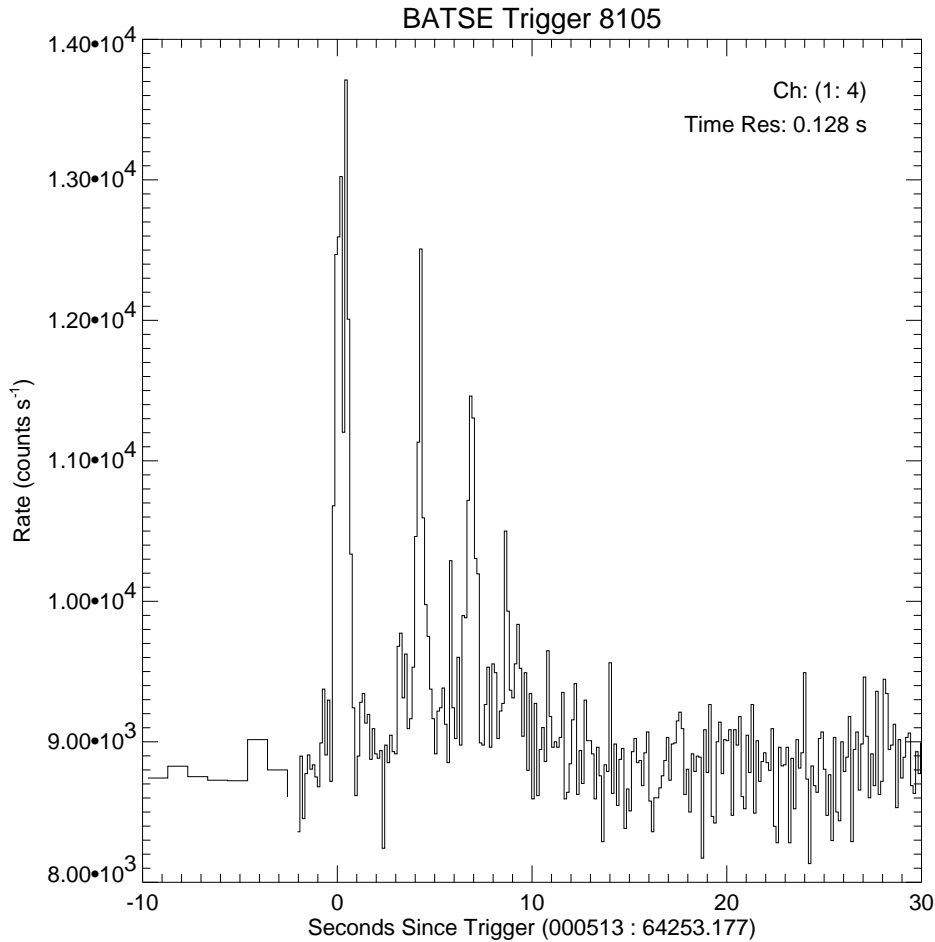}
\caption{An example of the temporal structure of a GRB as recorded by BATSE \cite{batse}. Note the two time scales:  a duration of several seconds and a fluctuation time scale of a fraction of a second.}
\label{seventeen}
\end{figure}

\subsubsection{A Brief History of Gamma Ray Bursts}

Gamma Ray Bursts (GRB) were accidentally discovered in the late 1960's by
the military Vela satellites, intended to monitor nuclear tests in
space forbidden by the Outer Space Treaty between the United States and
the Soviet Union \cite{hist1}.  The Vela observation of a short, intense
burst of MeV gamma rays was originally considered to be a possible
signal from an advanced extra-terrestrial civilization . The idea was
quickly reconsidered.  In 1973, the discovery was announced to the
public \cite{hist1}.  Shortly after, the observation was verified by the
Soviet IMP-6 satellite \cite{hist2}.

Until the 1990's, the high intensity of GRB led astronomers to the
belief that they were galactic in origin.  In 1991, the BATSE (Burst And
Transient Satellite Experiment) detector on the Compton Gamma Ray
Observatory was launched.  BATSE observed roughly one burst per
day within its field of view of about one third of the sky.  The
observations showed total isotropy of GRB over the entire sky, thus
ruling
out galactic origin \cite{hist3}.  The cosmological
origin of GRB implies that they release up to a solar mass of energy, in 
seconds time.  Their cosmological
origin was subsequently confirmed by afterglow observations, first made
in 1997 by the
Beppo-SAX satellite \cite{hist4}.  Afterglow observations were made in
X-ray, optical and longer wavelengths with an angular resolution of
arc-minute precision and with measurement of the redshift. To date,
dozens of GRB afterglows have been observed, nearly all of which have
resulted in the identification of the host galaxy 
\cite{host1,host2,host3}.

Although progress has been made in our understanding of GRB, many
questions remain unanswered.  Most importantly, the progenitor(s) of GRB
remain an open question.  In the next section, we describe some of the
most likely candidates.

\subsubsection{GRB Progenitors?}

The observed characteristics of GRB require an original event
with a large amount of energy ($\sim M_{\odot}$) in a very compact volume
($R_0\sim100\,$km). The phenomenology that describes observations is 
that of a fireball expanding with highly relativistic velocity, powered 
by radiation pressure. The nature of the ``inner engine'' that initiates 
the fireball remains an open question. Afterglow observations have 
recently shown that GRB are predominantly
generated in host galaxies and are likely the result of a stellar 
process.  Research into a variety of stellar progenitors has been 
pursued.

The ``Collapsar'' scenario, where a super-massive star undergoes core 
collapse resulting in a failed supernova, is one of the most common 
models proposed for the fireball's inner engine \cite{col1,col2}.  As 
matter falls into the black hole created in this process, gravitational 
energy is transfered to bulk kinetic energy and the fireball is 
generated.

The strength of magnetic fields and the angular momentum of the stellar 
object(s) involved can play an important role in the dynamics of the 
core collapse process. For example, ``Magnetars'' are a subset of the 
core collapse model which result in a rapidly spinning neutron star with 
an extremely strong magnetic field \cite{mag1,mag2,mag3}. Objects with 
sufficient angular momentum can undergo a ``supranova'' process where 
their core collapse takes place in two stages, possibly separated by 
months or years \cite{sup1,sup2}. In this scenario, the object's large 
angular momentum prevents a fraction of the matter from falling into the 
fireball initially.

Compact objects in close binary orbits are also likely candidates for 
fireball progenitors.  The ``hypernovae'' scenario is similar to the 
core collapse models, but includes a secondary stellar object in the 
dynamics \cite{pro1, pro8,pro9,pro10}.  Similarly, neutron star binaries 
or neutron star-black hole
binaries (or possibly white dwarf -- neutron star or black hole
binaries) which lose sufficient angular momentum through gravitation
radiation can undergo a merger.  Such a merger is expected to generate a
black hole surrounded by debris.  As this debris is accreted into the 
black hole, the required
fireball is generated \cite{pro2, pro3,pro4,pro5,pro6,pro7,pro9}.

Finally, if primordial strange hadrons exist, a ``seed'' of
strange matter may start a chain reaction converting a neutron star
into a
strange star made entirely of strange matter \cite{str1, str2,
str3}.  This conversion would release the majority of the star's binding
energy as it contracts, thus generating a compact fireball similar to
that
required for GRB dynamics.

Recent evidence indicates the presence of emission lines in GRB 
\cite{emission}. This evidence strengthens the argument for progenitors 
involving collapsing stars.

The problem of GRB progenitors is likely to have an experimental
solution.  Possible progenitor-specific signatures may be found by using
gravitational waves \cite{grav} or neutrinos as astronomical probes, or
by more detailed study of afterglows \cite{glow,glow2}.

It is interesting to note that the bimodal distribution of GRB durations
may be an indication of multiple GRB classes and associated progenitors.

\subsubsection{Fireball Dynamics}

\paragraph{The Fireball}

The dynamics of a gamma ray burst fireball is similar to the physics of
the early universe.  Initially, there is a radiation dominated soup of
leptons and photons and few baryons. This perfect fluid has the equation
of state
$P=\rho/3$ and is initially hot enough to freely produce
electron-positron
pairs.  The luminosity of a burst can be related to the number density of
photons $n_\gamma$:
\begin{eqnarray}
L = 4 \pi R_0^2 c n_\gamma E_\gamma,
\end{eqnarray}
where $R_0$ is the initial radius of the source, i.e. prior to
expansion.  The
optical depth of a photon before pair production is determined by the
photon density and the interaction cross section \cite{waxman1}:
\begin{eqnarray}
\tau_{\rm{opt}} = \frac{R_0}{\lambda_{\rm{int}}} = R_0 n_\gamma 
\sigma_{\rm{Th}}
=\frac{L\sigma_{\rm{Th}}}{4\pi R_0
c E_\gamma}\sim 10^{15}\bigg(\frac{L_{\gamma}}{10^{52}
\rm{erg/s}}\bigg)\bigg(\frac{100 \,\rm{km}}{R_0}\bigg)\bigg(\frac{1\,
\rm{MeV}}{E_\gamma}\bigg).
\end{eqnarray}
Here $\lambda_{\rm{int}}$ is the interaction length of a photon as a
result of pair production and Thomson scattering. These cross sections
are roughly equal with the Thomson cross section
$\sigma_{\rm{Th }}\simeq 10^{-24}\rm{cm}^2$.

With an optical depth of order $\sim 10^{15}$, photons are trapped
in the fireball.  This results in the highly relativistic expansion of
the fireball powered by
radiation pressure \cite{fire1,pro7}. The fireball will expand with
increasing velocity until it becomes transparent and the radiation is
released. This results in the visual display of the GRB. By this time,
the expansion velocity has reached highly relativistic values of order
$\gamma\simeq300$.

Besides leptons and photons, the fireball contains some baryons. During
expansion, the opaque fireball cannot radiate and any nucleons present
are accelerated as radiation is converted into bulk kinetic energy. When 
the radiation is emitted, there is a transition from
radiation to matter dominance of the fireball. At this stage, the 
radiation pressure is no longer
important and the expanding fireball coasts without acceleration.  The 
expansion velocity remains constant with $\gamma\simeq \eta
\equiv\frac{L}{\dot{M}c^2}$ that  is determined by the amount of
baryonic matter present, often referred to as the baryon loading 
\cite{fire2,fire3}.
\begin{figure}[thb]
\centering\leavevmode
\includegraphics[width=5in]{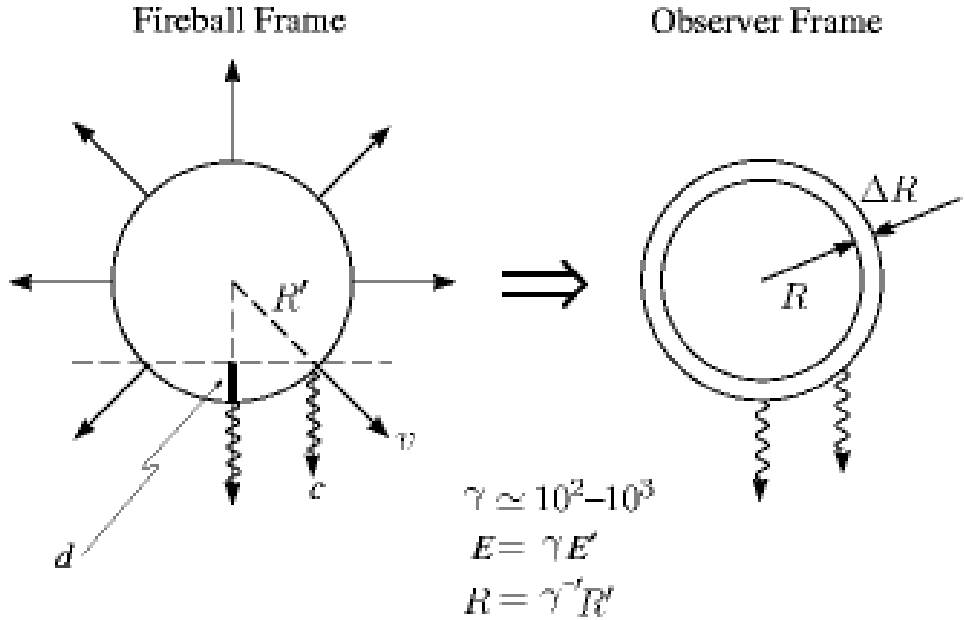}
\caption{Diagram of GRB fireball kinematics assuming no beaming. Primed
quantities refer to the comoving frame. Unprimed quantities refer to the
observer's frame.}
\label{seventeenb}
\end{figure}
The phenomenology will reveal values of $\eta$ between $10^2$ and $10^3$ 
\cite{waxman1,waxman2,piran}. The formidable appearance of the GRB
display is simply associated with the large boost between the fireball
and the observer who detects highly boosted energies and contracted
times.

The exploding fireball's original size, $R_0$, is that of the compact 
progenitor, for instance the black hole created by the collapse of a 
massive star. As the fireball expands the flow is shocked in ways 
familiar from the emission of jets by the black holes at the centers of 
active galaxies or mini-quasars. (A way to visualize the formation of 
shocks is to imagine that infalling material accumulates and chokes the 
black hole. At this point a blob of plasma is ejected. Between these 
ejections the emission is reduced.) The net result is that the expanding 
fireball is made up of multiple shocks. These are the sites
of the acceleration of particles to high-energy and the seeds for the
complex millisecond structures observed in individual bursts; see 
Fig.\,16. (Note that these shocks expand with a range of velocities and 
they will therefore collide providing a mechanism to accelerate 
particles to high-energy.) The
characteristic width of these shocks in the fireball frame is
$\delta R^\prime=\gamma c \Delta t$, where $\Delta t \simeq 0.01\,$sec. 
For an alternative scenario; see Ref.\,\cite{cannon1, cannon2, cannon3}.

An expanding shock is seen by the observer as an expanding shell of 
thickness
$c\Delta t=R_0$ and radius $R$; see Fig.\,17. Here $\Delta t$ is the 
time scale of fluctuations in the burst fireball; it is related to 
$R^\prime$ by:
\begin{eqnarray}
R^\prime = \gamma^2 c \Delta t = \gamma^2 R_0,
\end{eqnarray}
with primed quantities referring from now on to the frame where the
fireball is at rest. Two, rather than a single $\gamma$-factor, 
relate
the two quantities because of the geometry that relates the radius
$R^\prime$ to the time difference between  photons emitted from a shell
expanding
with a velocity $v$; see Fig.\,17 \cite{halzenlec}. Introducing the
separation $d$ of the two
photons along the line of sight, we note that
\begin{eqnarray}
\Delta t \sim\frac{d}{c}\sim\frac{1}{c}(R^\prime-
R^\prime\frac{v}{c})\simeq
\frac{R^\prime}{2c}(1-\frac{v^2}{c^2})\sim\frac{R^\prime}{c\gamma^2},
\end{eqnarray}
using the relativistic approximation that $1+\frac{v^2}{c^2}\simeq1$.

We next calculate the energy of the burst. In the observer frame
\begin{eqnarray}
E=U \times V \propto T^4 \times R^2 \Delta R \propto T^4 R^2 \propto
\gamma^4 T^{\prime\,4} R^2,
\end{eqnarray}
where $U$ is the energy density and $V$ is the fireball shell volume.  In the fireball frame,
\begin{eqnarray}
E^{\prime} = U^{\prime} \times V^{\prime} \propto T^{\prime\,4}
R^{\prime\,3} \propto \gamma^{6} T^{\prime\,4}.
\end{eqnarray}
Energy conservation requires that E and $ E^{\prime}$ remain constant
during expansion of the fireball. In the fireball frame, this results
in the usual blackbody relation that
$T^{\prime\,4}$ is proportional to $R^{\prime\,-3}$ or, using Eq.\,18,
proportional to $\gamma^{-6}$. Substituting into the expression for E, we
obtain that
\begin{eqnarray}
E \propto \gamma^{-2}\,R^2,
\end{eqnarray}
or, because E is constant, that
\begin{eqnarray}
R\propto \gamma.
\end{eqnarray}
Thus we obtain the important result that, with expansion, the
$\gamma$-factor grows linearly with R until reaching the maximum value 
$\eta$.

\paragraph{The Observed GRB Spectrum: Synchroton and Inverse Compton
Scattering}

The broken power-law gamma ray spectrum of GRB, with two distinct
spectral slopes, is far from a blackbody spectrum. The observed spectrum,
therefore, clearly indicates that fireball photons do not sufficiently
interact to thermalize prior to escaping the fireball. After escaping, 
the photons show spectral features characteristic of the high-energy,
non-thermal emission by supernova remnants and active galaxies. Here
photons up to MeV-energy can be produced by synchrotron radiation, with
some reaching, possibly, up to TeV energies by inverse Compton 
scattering on
accelerated electrons \cite{piran,milagro}. These processes have been modeled for 
the
expanding fireball and successfully accommodate observed
GRB spectra.  This represents a major success of the
relativistic fireball phenomenology \cite{syn1,syn2,syn3,syn4,syn5}.

To produce the non-thermal spectrum, special conditions must
prevail \cite{waxman1,waxman2,halzenlec,piran}. The photons must not
thermalize prior to the time when the shock becomes transparent and the
observed radiation released. Conversely, if they decouple too early,
there is insufficient time for the synchotron and inverse Compton
scattering processes to produce the observed spectrum. This requires
that the expansion time of the shockwave and the time 
$\lambda_{\rm{int}}/c$  for photons and electrons to interact by 
Thompson scattering be
similar:
\begin{eqnarray}
\frac{R}{\gamma c} \sim (n_e c \sigma_T)^{-1}.
\end{eqnarray}
Here $\sigma_T$ is the Thompson cross section and $n_e$ is the electron 
number density in the fireball. The latter can be
related to the mass flux $\dot{M}$ with the assumption that $n_e
\cong n_p$:
\begin{eqnarray}
\dot{M} = 4 \pi R^2 c \gamma n_p m_p.
\end{eqnarray}
As required by mass conservation, $\dot{M}$ is independent of R since
$n_p\propto \gamma^{-3}$ and $R\propto \gamma$. In terms of luminosity,
\begin{eqnarray}
n_e \cong \frac{\dot{M}}{4 \pi R^2 c \gamma m_p}  =   \frac{L}{\gamma
\eta
c^3 4 \pi R^2 m_p},
\end{eqnarray}
where $\eta = L/\dot{M}c^2$ is the ratio of luminosity to mass 
previously introduced. $\eta$ is also referred to as the dimensionless
entropy and, as previously derived, $\gamma \simeq \eta$ after expansion 
of the fireball. We can rewrite the condition of Eq.\,21 for producing 
the
observed non-thermal spectrum as
\begin{eqnarray}
\frac{R n_e \sigma_T}{\gamma} = \frac{L \sigma_T}{\gamma^2 \eta c^3 4 \pi
R m_p}= \frac{1}{\gamma^3} \frac{\eta^{*4}}{\eta} \simeq 1,
\end{eqnarray}
where the critical dimensionless entropy $\eta^*$ is defined as
\begin{eqnarray}
\eta^* \equiv \bigg(\frac{L \sigma_T}{c^3 4 \pi R_0 m_p}\bigg)^{1/4}
\simeq 1000 \times \bigg(\frac{L}{10^{52} \rm{erg/s}} \bigg)^{1/4}
\bigg(\frac{100\, \rm{km}}{R_0} \bigg)^{1/4}.
\end{eqnarray}
At decoupling, $\eta  = \gamma$ and Eq.\,24 is satisfied, provided
$\eta = \eta^*$. The condition for the fireball to produce the correct
non-thermal, synchrotron/inverse Compton spectrum is realized with the expansion time matching the Thompson scattering time. For values of $\eta$ that
are significantly larger (smaller), the decoupling of the radiation will
occur too early (late) thus limiting $\eta$ , as well as the final value
of $\gamma$, to the range of $10^2-10^3$.

\paragraph{Jets and Beaming}

Observations imply that the total amount of energy emitted in gamma rays
by a GRB are typically in the range of $10^{52}-10^{54}$ ergs, i.e. a
large fraction of a solar mass. For some bursts, it may exceed a solar
mass. This, as well as the difficulty of converting such an unusually
large fraction of primary energy into gamma rays, strongly suggests that
GRB are beamed. Beaming reduces the total amount of energy by a factor 
of $\Omega / 4\pi$, where $\Omega$ is the solid angle in which the 
observed
gamma rays are
emitted. Most proposed progenitors naturally predict a
rotating
stellar source that is likely to produce beamed emission.  Relativistic
beaming is possible down to an angular size of
$\Omega > \gamma^{-2}$, although larger angles are of course
possible \cite{beam1, beam2, beam3}.

In the presence of beaming, the number of bursts is increased
by a factor of $4\pi/\Omega$ in order to account for bursts that do not
point towards earth and are, therefore, not observed.  For typical 
Lorentz
factors of
$\gamma\sim300$ and a minimum beaming angle of $\Omega > \gamma^
{-2}\sim10^{-5}$, on the
order of one GRB per galaxy per year is required to accommodate the
observations \cite{beam2}.  It is important to note that most of the
diffuse neutrino fluxes calculated in this review are independent of 
beaming because the
reduced energy for a single burst is compensated by their increased
frequency.

\subsubsection{Ultra High-energy Protons From GRB?}

As previously discussed, it may be possible to accelerate protons to
energies above $10^{20}$\,eV in GRB shocks \cite{waxman2,waxman2b,waxman2c}. GRB
within the GZK radius of
50-100 Mpc, could therefore be the source of the ultra high-energy
cosmic rays
(UHECR's) \cite{dermer, waxman2,waxman2b,waxman2c, waxman3,
uhecr1,uhecr2,uhecr3,mostlum3,olinto}.  To accelerate protons to this
energy, several conditions have to be satisfied.  First, the
acceleration time
$t_a \sim A R_L/c$, where A is a factor of order 1 and $R_L=E/eB\gamma$
is the Larmor radius, must not exceed the duration of the burst
$R/\gamma c$,
\begin{eqnarray}
\frac{A E}{\gamma e B} \lsim \frac{R}{\gamma},
\end{eqnarray}
or
\begin{eqnarray}
B \gsim\frac{A E}{e R} \simeq A \times 10 \,\rm{tesla}\,
\bigg(\frac{E}{10^{20}\rm{eV}}\bigg) \bigg(\frac{10^{11}\rm{m}}{R}\bigg).
\end{eqnarray}
Second, energy losses due to synchrotron radiation must not exceed the
energy gained by acceleration. The synchrotron loss time is given by
\begin{eqnarray}
t_{syn}=\frac{\lambda_{\rm{int}}}{c}=\frac{1}{c n_e \sigma_T}.
\end{eqnarray}
The number density of electrons (in the rest frame) is given by
\begin{eqnarray}
n_e=\frac{m^2_e c^4 B^2}{6 \pi}.
\end{eqnarray}
Therefore,
\begin{eqnarray}
t_{syn}=\frac{6 \pi }{\sigma_T m_e^2 c^5 B^2 \gamma_p}=\frac{6 \pi 
m_p^4c^3}{\sigma_T m_e^2 E B^2}.
\end{eqnarray}
For synchrotron energy losses to be less than the energy gained by 
acceleration,
\begin{eqnarray}
\frac{6 \pi m_p^4c^3}{\sigma_T m_e^2 E B^2} \gsim A\times
\frac{E}{\gamma e c B},
\end{eqnarray}
or
\begin{eqnarray}
B \lsim \frac{1}{A} \times 10 \,
\rm{ tesla} \bigg(\frac{\gamma}{300}\bigg)^2
\bigg(\frac{10^{20}\rm{eV}}{E}\bigg)^2.
\end{eqnarray}
Combining above requirements, we get
\begin{eqnarray}
     A \times 10
\bigg(\frac{E}{10^{20}\rm{eV}}\bigg) \bigg(\frac{10^{11}\rm{m}}{R}\bigg)
\rm{
tesla} \lsim B \lsim  \frac{1}{A} \times 10 \, \rm{ tesla}
\bigg(\frac{\gamma}{300}\bigg)^2 \bigg(\frac{10^{20}\rm{eV}}{E}\bigg)^2,
\end{eqnarray}
or
\begin{eqnarray}
R \gsim A^2 \times 10^{10} \, \rm{ meters}
\bigg(\frac{300}{\gamma}\bigg)^2 \bigg(\frac{E}{10^{20}\rm{eV}}\bigg)^3.
\end{eqnarray}
 From simple fireball kinematics, we previously derived that
\begin{eqnarray}
R \lsim \gamma^2 c \Delta t,
\end{eqnarray}
where $\Delta t$ is $\sim10$ msec. Combining
this with Eq.\,34 leads to the final requirement:
\begin{eqnarray}
A^2 \times 10^{10} \bigg(\frac{300}{\gamma}\bigg)^2
\bigg(\frac{E}{10^{20}\rm{eV}}\bigg)^3 \lsim R \lsim \gamma^2 c \Delta t,
\end{eqnarray}
or
\begin{eqnarray}
\gamma \gsim A^{1/2}\times 130
\bigg(\frac{E}{10^{20}\rm{eV}}\bigg)^{3/4} \bigg(\frac{.01
\rm{sec}}{\Delta t}\bigg)^{1/4},
\end{eqnarray}
which can indeed be satisfied for the values of $\gamma=10^2-10^3$
previously derived from fireball phenomenology. We conclude that bursts 
with Lorentz factors $\gsim 100$ can accelerate
protons to $\sim 10^{20}$ eV. The long acceleration time of
10-100 seconds implies however that the fireball extends to a large
radius where surrounding matter may play an important role in the
kinematics of the expanding shell.

Finally, it can be shown that proton energy losses from $p-\gamma$
interactions
will not interfere with acceleration to high-energy. These will, in fact,
be the source
of high-energy neutrinos associated with the beam of high-energy
protons. We will discuss this further on.

\subsubsection{Neutrino Production in GRB: the Many Opportunities}

Several mechanisms have been proposed for the production of neutrinos in
GRB. We summarize them first:
\begin{itemize}

\item{}Thermal Neutrinos:  MeV Neutrinos \\
As with supernovae, GRB are expected to radiate the vast majority of
their initial energy as thermal neutrinos.  Although the details are
complex and are likely to depend on the progenitor, a neutrino spectrum
with a higher temperature than a supernova may be expected. Observation 
is
difficult because of the great distances to GRB, although we should keep 
in mind
that a nearby GRB may not be less frequent than a galactic supernova;
see the section on beaming.

\item{}Shocked Protons: TeV-EeV Neutrinos \\
Protons accelerated in GRB can interact with fireball gamma rays
and produce pions that decay into neutrinos. While astronomical
observations provide information on the fireball gamma rays, the proton
flux is a matter of speculation. A definite neutrino flux is however
predicted when assuming that GRB produce the highest energy cosmic rays.

\item{}Decoupled Neutrons:  GeV Neutrinos \\
In a GRB fireball, neutrons can decouple from protons in the expanding 
fireball. If their relative velocity is sufficiently high, their
interactions will be the source of pions and, therefore, neutrinos. 
Typical energies
of the neutrinos produced are much lower than those resulting
from interactions with gamma rays.

\end{itemize}

\subsubsection{Thermal MeV Neutrinos from GRB}

As is the case for a supernova, we expect that in GRB, thermal
neutrinos are produced escaping with the majority of the total energy 
\cite{thermal1, thermal2}. The dynamics are somewhat different, however. The temperature of the photons in a supernova
is of order 10\,MeV as derived from the familiar estimate: %
\begin{eqnarray}
U_{\gamma}=\frac{4 \sigma T^4}{c}\rightarrow T_\gamma \simeq
\bigg(\frac{E_{\rm{in}\,\gamma \rm{'s}}}{V}\frac{c}{4 \sigma}\bigg)^{1/4}
\simeq 11\, \rm{MeV} \bigg(\frac{E_{\rm{in}\,\gamma
\rm{'s}}}{10^{52}\rm{ergs}}\bigg)^{1/4} \bigg(\frac{{100}\,
\rm{km}}{R}\bigg)^{3/4},
\end{eqnarray}
where $\sigma\simeq 5.67 \times 10^{-8} \rm{km}/\rm{sec}^3$ is the
Stefan-Boltzman constant and $U_{\gamma}$ is the energy density of the
initial plasma. We copy this estimate for the neutrinos produced in the
pre-fireball phase of a GRB taking into account the much larger energy
emitted, typically 2 orders of magnitude:
\begin{eqnarray}
T_\nu = T_\gamma \bigg(\frac{E_{\rm{in}\,\nu \rm{'s}}/E_{\rm{in}\,\gamma
\rm{'s}}}{h_\nu/h_\gamma}\bigg)^{1/4} \simeq 28 \, \rm{MeV}
\bigg(\frac{E_{\rm{in}\,\nu \rm{'s}}}{10^{54}\rm{ergs}}\bigg)^{1/4}
\bigg(\frac{{100}\, \rm{km}}{R}\bigg)^{3/4}.
\end{eqnarray}
Here $h_\nu=2\times 3\times 7/8=21/4$ and $h_\gamma=2$ are the degrees
of freedom available to each particle type. This yields neutrinos
with average energy:
\begin{eqnarray}
E_{\nu\rm{,ave}} \simeq 3.15 \, T_\nu \simeq 90 \, \rm{MeV}
\bigg(\frac{E_{\rm{in}\,\nu \rm{'s}}}{10^{54}\rm{ergs}}\bigg)^{1/4}
\bigg(\frac{{100}\, \rm{km}}{R}\bigg)^{3/4}.
\end{eqnarray}
The flux of neutrinos can now be calculated from the average energy of
roughly 100\,MeV per individual neutrino and the total energy
available:
\begin{eqnarray}
N_\nu \simeq \frac{E_{\rm{in}\,\nu \rm{'s}}}{E_{\nu\rm{,ave}}}
\frac{1}{4\pi D^2} \simeq 6\times 10^{10}\, \rm{km}^{-2} \,
\bigg(\frac{E_{\rm{in}\,\nu \rm{'s}}}{10^{54}\rm{ergs}}\bigg)^{3/4}
\bigg(\frac{R}{100\, \rm{km}}\bigg)^{3/4}  \bigg(\frac{3000 \,
\rm{Mpc}}{D}\bigg)^{2}.
\end{eqnarray}
Neutrinos with this energy are below threshold for the detection methods
previously
described. For supernova 1987A, predominantly electron anti-neutrinos
were observed by the electromagnetic showers generated inside the
detector by the process $\bar{\nu_e} + p\rightarrow n + e^+$.
Underground detectors are too small to detect the above flux because of 
the
cosmological distance to the source. In large under-ocean detectors, the
signal is drowned in the $\sim50\,$kHz noise in the photomultipliers from
potassium decay. Only a large Cerenkov detector embedded in
sterile ice can possibly detect GRB but
even here the signal-to-noise is marginal unless the source is within
our local cluster. Detailed calculations have been performed in
Ref.\, \cite{thermal1}.

\subsubsection{Shocked Protons: PeV Neutrinos}

Assuming that GRB are the sources of the highest energy cosmic rays and
that the efficiency for conversion of fireball energy into the kinetic
energy of protons is similar to that for electrons, the production of
PeV neutrinos is a robust prediction of the
relativistic fireball model \cite{waxmanbahcall, halzenlec, dermer}.
Neutrinos are produced in interactions of accelerated protons with
fireball photons, predominantly via the processes
\begin{eqnarray}
p\gamma \rightarrow \Delta \rightarrow n \pi^{\pm}
\end{eqnarray}
and
\begin{eqnarray}
p\gamma \rightarrow \Delta \rightarrow p \pi^{0}
\end{eqnarray}
which have very large cross sections of $10^{-28}
\rm{cm}^2$.  The charged $\pi$'s subsequently decay producing charged
leptons and
neutrinos, while neutral $\pi$'s may generate high-energy photons
observable in TeV energy air Cerenkov detectors.

For the center-of-mass energy of a proton-photon interaction to exceed
the threshold energy for producing the $\Delta$-resonance, the comoving
proton energy must exceed
\begin{eqnarray}
E^\prime_p > \frac{m_\Delta^2-m_p^2}{4 E^\prime_\gamma}.
\end{eqnarray}
Therefore, in the observer's frame,
\begin{eqnarray}
E_p > 1.4 \times 10^{16} \rm{eV} \bigg(\frac{\gamma}{300}\bigg)^2
\bigg(\frac{1 MeV}{E_\gamma}\bigg),
\end{eqnarray}
resulting in a neutrino energy
\begin{eqnarray}
E_\nu = \frac{1}{4} \langle x_{p \rightarrow \pi} \rangle E_p > 7 \times
10^{14} \rm{eV} \bigg(\frac{\gamma}{300}\bigg)^2 \bigg(\frac{1
MeV}{E_\gamma}\bigg).
\end{eqnarray}
Here $\langle x_{p \rightarrow \pi} \rangle \simeq .2$ is the average
fraction of energy transferred from the initial proton to the
produced pion.  The factor of 1/4 is based on the estimate that the 4
final state leptons in the decay chain
$\pi^{\pm} \rightarrow \bar{\nu_{\mu}} \mu \rightarrow\bar{\nu_{\mu}} e 
\nu_e \bar{\nu_e}$ equally share the pion energy.

We already discussed the fireball's expansion and the formation of 
shocks that are the sites
of the acceleration of particles to high-energy and the seeds for the
complex millisecond structures observed in individual bursts. The
characteristic width of a shock in the fireball frame is
$\Delta R^\prime=\gamma c \Delta t$, where $\Delta t \simeq 0.01\,$sec.

As the kinetic energy in fireball protons increases with expansion, a
fraction of this energy is converted into pions once the protons are
accelerated above threshold for pion production. The
fraction of energy converted to pions is estimated from the number of
proton interactions occurring within a shock of characteristic
size $\Delta R^\prime$:
\begin{eqnarray}
f_{\pi} \simeq \frac{\Delta R^\prime}{\lambda_{p \gamma}}\langle
x_{p\rightarrow \pi}\rangle.
\end{eqnarray}
The proton interaction length $\lambda_{p \gamma}$ in the photon
fireball is given by
\begin{eqnarray}
\frac{1}{\lambda_{p\gamma}}=n_\gamma \sigma_{\Delta}.
\end{eqnarray}
Here $n_\gamma$ is the number density of photons in the fireball frame
and
$\sigma_\Delta \sim 10^{-28} \rm{cm}^2$ is the proton-photon cross
section at the $\Delta$-resonance.  The photon number density is the
ratio of the photon energy density and the photon energy in the comoving
frame:
\begin{eqnarray}
n_\gamma=\frac{U_\gamma^\prime}{E_{\gamma}^\prime}=\bigg(\frac{L_\gamma
\Delta t/\gamma}{4\pi R^{\prime 2} \Delta R^\prime}\bigg)   \bigg/
\bigg(\frac{E_\gamma}{\gamma}\bigg).
\end{eqnarray}
Using the fireball kinematics of Eqs.\,15 and 16
\begin{eqnarray}
n_\gamma=\bigg(\frac{L_\gamma}{4\pi c^3 \Delta t^2
\gamma^6}\bigg)   \bigg/
\bigg(\frac{E_\gamma}{\gamma}\bigg)=\frac{L_\gamma}{4\pi c^3 \Delta t^2
\gamma^5 E_{\gamma}}.
\end{eqnarray}
Thus we obtain the fraction of proton energy converted to $\pi$'s in the
expansion:
\begin{eqnarray}
f_{\pi} \simeq \frac{L_{\gamma}}{E_{\gamma}}\frac{1}{\gamma^4 \Delta t}
\frac{\sigma_\Delta \langle x_{p \rightarrow \pi} \rangle}{4 \pi c^2}
\simeq .13 \times
\bigg(\frac{L_\gamma}{10^{52}\rm{erg/s}}\bigg) \bigg(\frac{1
\rm{MeV}}{E_\gamma}\bigg)  \bigg(\frac{300}{\gamma}\bigg)^4
\bigg(\frac{.01 \rm{sec}}{\Delta{t}}\bigg).
\end{eqnarray}
For $L \sim 10^{52}$ erg/sec, $\Delta t \sim 10$ msec and $\gamma \simeq$
300, this fraction is on the order of 10 percent. This quantity strongly
depends on the Lorentz factor $\gamma$.  Even modest
burst-to-burst fluctuations in $\gamma$ around the average value of 300
can result in a PeV neutrino flux dominated by a few bright
bursts; we will return to a discussion of fluctuations further
on \cite{fluc1, fluc2}.

In order to normalize the neutrino flux we introduce the assumption that
GRB are the source of cosmic rays above the ankle of the cosmic ray
spectrum near $\sim 3 \times 10^{18}$
eV \cite{dermer, waxman2,waxman2b,waxman2c,
waxman3,uhecr1,uhecr2,uhecr3,mostlum3}.  The flux in neutrinos can then be
simply obtained from the total energy injected into cosmic rays and the
average energy of a single neutrino \cite{waxmanbahcall}:
\begin{eqnarray}
\phi_\nu \simeq
\frac{c}{4\pi}\frac{U^\prime_\nu}{E^\prime_\nu}=\frac{c}{4
\pi}\frac{U_\nu}{E_\nu}=\frac{c}{4 \pi}\frac{1}{E_\nu}\bigg( \frac{1}{2}
f_\pi t_H \frac{dE}{dt} \bigg),
\end{eqnarray}
or
\begin{eqnarray}
\phi_\nu = 2 \times 10^{-14} \rm{cm}^{-2} \rm{s}^{-1} \rm{sr}^{-1} \bigg(
\frac{7 \times 10^{14}\rm{eV}}{E_\nu} \bigg) \bigg(\frac{f_\pi}{0.125}
\bigg) \bigg(\frac{t_H}{10 \rm{Gyr}} \bigg) \bigg( \frac{dE/dt}{4 \times
10^{44}\, \rm{Mpc}^{-3} \rm{yr}^{-1}} \bigg),
\end{eqnarray}
where $t_H \sim 10$ Gyrs is the Hubble time and $dE/dt \sim 4 \times
10^{44}\, \rm{ergs}\,\rm{Mpc}^{-3}\,\rm{yr}^{-1}$ is the injection rate
of energy into the universe in the form of cosmic rays above the ankle.

PeV neutrinos are detected by observing a charged lepton produced in the
charged current interaction of a neutrino near the detector.  For
instance, the
probability to detect a muon neutrino within a Cerenkov neutrino
telescope's effective area is given by Eq.\,11. At TeV-PeV energies the
function $ P_{\nu \rightarrow \mu}$ can be approximated by
\begin{eqnarray}
P_{\nu \rightarrow \mu} \simeq  1.7 \times 10^{-6}
E^{0.8}_{\nu,obs}(\rm{TeV}),
\end{eqnarray}
where $E_{\nu,obs}=E_{\nu}/(1+z)$ is the observed neutrino energy.  The
rate of detected events is the convolution of the flux with the
probability of detecting the neutrino
\begin{eqnarray}
N_{events} = \int_{E_{\rm{thresh}}}^{E_\nu^{\rm{max}}} \phi_\nu P_{\nu
\rightarrow \mu} \frac{dE_\nu}{E_\nu} \simeq 25 \,\rm{km}^
{-2}\rm{yr}^{-1}
\end{eqnarray}
This rate is significantly enhanced when fluctuations in distance, 
energy and (possibly) Lorentz factor are considered.  The event rate is
likely
to be dominated by a few bright bursts rather than by a diffuse flux.

With the ability to look for GRB neutrino events in coincidence with
gamma ray observations, i.e. in short time windows over which very
little background accumulates, there is effectively no background for
this neutrino signature of GRB.

\subsubsection{Stellar Core Collapse: Early TeV Neutrinos}

The core collapse of massive stars is, arguably, the most promising
mechanism for generating GRB. The fireball produced is likely to be
beamed in jets along the collapsed object's rotation axis. The mechanism
is familiar from observations of jets associated with the central black
hole in active galaxies. The jets subsequently run into the stellar
matter accreting onto the black hole. If the jets successfully emerge
from the stellar envelope a GRB results. Interestingly, failed
``invisible'' jets which
do not emerge will not produce a GRB display but will still produce
observable neutrinos \cite{waxmantev1,waxmantev2}.

A beamed GRB jet expanding with a Lorentz factor $\gamma_{\rm{jet}}\sim
100-1000$
through the stellar envelope, will be slowed down resulting in a smaller
Lorentz factor at its leading edge $\gamma_{\rm{f}} \ll
\gamma_{\rm{jet}}$.  Therefore, the fast particles in the tail will catch
up with the slow particles in the leading edge and collide with a
Lorentz factor $\gamma\approx\gamma_{\rm{jet}}/2\gamma_{\rm{f}}$ by 
simple
addition of relativistic velocities. Once the jet emerges from the
infalling stellar matter  around $R\sim 10^6$ km, the density drops to
around $\sim 10^{-7} \rm{g}/\rm{cm}^3$.  Matching the
energy densities on either side of the shock front requires
\begin{eqnarray}
\gamma_{\rm{f}}^2 \times 10^{-7} \rm{g}/\rm{cm}^3 \simeq
\bigg(\gamma_{\rm{jet}}/2\gamma_{\rm{f}}\bigg)^2 n_p m_p,
\end{eqnarray}
where $n_p$ is the comoving proton number density in the jet. It is
related to the luminosity of the burst; see Eq.\,23 :
\begin{eqnarray}
n_p=\frac{L_{\rm{iso}}}{4\pi R^2 \gamma_{\rm{jet}}^2 m_p c^3}.
\end{eqnarray}
Here $L_{\rm{iso}}=L \frac{\Omega}{4\pi}$ is the inferred from a 
non-beamed flux. Eqs.\,56 and 57 determine the
value of $\gamma_{\rm{f}}$:
\begin{eqnarray}
\gamma_{\rm{f}}\simeq \bigg(\frac{L_{\rm{iso}}}{16 \pi \rho R^2
c^3}\bigg)^{1/4}\simeq 3\times
\bigg(\frac{L_{\rm{iso}}}{10^{52}\rm{erg/s}}\bigg)^{1/4}
\bigg(\frac{10^{-7}\rm{g}/\rm{cm}^3}{\rho}\bigg)^{1/4}
\bigg(\frac{10^{7}\rm{km}}{R}\bigg)^{1/2}.
\end{eqnarray}
With this low value of the Lorentz factor, the fireball remains opaque
to gamma rays as described in the section on fireball dynamics. The
radiation thermalizes with a temperature determined by its energy
density $U^\prime_{\gamma}$ and the Stefan-Boltzman law 
$U^\prime_{\gamma}= 4 \sigma
T^{\prime 4}/c$. We find

\begin{eqnarray}
T^\prime_\gamma \simeq \bigg(\frac{4
\gamma_{\rm{f}}^2 \rho c^3}{4 \sigma}\bigg)^{1/4} \simeq 2.2 \,\rm{keV}
\bigg(\frac{\rho}{10^{-7}\rm{g}/\rm{cm}^3}\bigg)^{1/4}
\bigg(\frac{\gamma_{\rm{f}}}{3}\bigg)^{1/2}.
\end{eqnarray}
The rest of the calculation follows the previous section. Protons
traveling
through this thermal photon plasma produce pions, predominantly via the
$\Delta$-resonance,  for energies
\begin{eqnarray}
E^\prime_p > \frac{m_\Delta^2-m_p^2}{4 E^\prime_\gamma} \simeq 7\times
10^4 \,\rm{GeV} \bigg(\frac{2.2\, \rm{keV}}{E^\prime_\gamma}\bigg).
\end{eqnarray}
Primed energies refer to the comoving frame with
$\gamma_{\rm{f}}E^\prime=E$.  Neutrinos emerge from the interactions
with energy
\begin{eqnarray}
E_\nu = \frac{1}{4} \langle x_{p \rightarrow \pi} \rangle E_p \simeq 10
\,\rm{TeV} \bigg(\frac{\gamma_{\rm{f}}}{3}\bigg) \bigg(\frac{2.2\,
\rm{keV}}{T^\prime_\gamma}\bigg).
\end{eqnarray}

For mildly relativisitic conditions, the fraction of protons converted to
pions is expected to be high in the very dense plasma, i.e. of order
unity. The neutrino flux observed from a single GRB at a distance D is
calculated from the total energy emitted in neutrinos and the average
energy of a single neutrino:
\begin{eqnarray}
\phi_\nu \simeq \frac{E_{\rm{iso}} \langle x_{p \rightarrow \pi}
\rangle}{16 \pi D^2 E_\nu} \simeq .003 \, \nu '{\rm{s}} \,\, 
{\rm{m}}^{-2} \times \bigg( \frac{E_{\rm{iso}}}{10^{53}\rm{ergs}}\bigg) 
\bigg(\frac{10
\rm{TeV}}{E_\nu}\bigg) \bigg(\frac{3000 \rm{Mpc} } { D }\bigg)^2.
\end{eqnarray}
This flux of TeV neutrinos from a single burst results in
\begin{eqnarray}
N_{events} \sim \phi_\nu \times P_{\nu \rightarrow \mu}\sim .05
\bigg(\frac{E_{\rm{iso}}}{10^{53}\rm{ergs}}\bigg) \bigg(\frac{3000
\rm{Mpc}}{D}\bigg)^2
\end{eqnarray}
events observed per year, per square kilometer of the detector. Here
$E_{\nu,obs}=E_{\nu}/(1+z)$ is the observed
neutrino energy and, for TeV-energy neutrinos, we used the approximation
\begin{eqnarray}
P_{\nu \rightarrow \mu} \simeq  1.3 \times 10^{-6} E_{\nu,obs}(\rm{TeV}).
\end{eqnarray}

The event rate is low. Bursts within a few hundred megaparsecs ($\sim
10$ bursts per
year as well as an additional unknown number of ``invisible'' bursts from
failed GRB) may produce multiple TeV neutrino events in a kilometer scale
detector.  This signature is unique to supernova progenitors.

\subsubsection{UHE Protons From GRB: EeV Neutrinos}

Recent observations of GRB afterglows show evidence that GRB explode
into an interstellar medium, consistent with the speculations that they
are collapsing or merging stars. Shocks will be produced when the GRB
runs into the interstellar medium, including a reverse shock that
propagates back into the burst ejecta.  Electrons and
positrons in the reverse shock radiate an afterglow of eV-keV photons
that represent a target
for neutrino production by ultra high-energy protons accelerated in the
burst \cite{waxman10}.

The fraction of proton energy going into $\pi$-production is calculated
as before following Eq.\,47,
\begin{eqnarray}
f_{\pi} \simeq \frac{\Delta R^\prime}{\lambda_{p \gamma}}\langle
x_{p\rightarrow \pi} \rangle,
\end{eqnarray}
\begin{eqnarray}
f_{\pi} \simeq \frac{L_{\gamma}(E_{\gamma, min})}{E_{\gamma, min}}
\frac{1}{\gamma_{rs}^4 \Delta t} \frac{\sigma_\Delta \langle x_{p
\rightarrow \pi} \rangle}{4 \pi c^2},
\end{eqnarray}
where $\gamma_{rs}$ is the Lorentz factor of the reverse shock. For the
afterglow, the relevant time scale is 10-100 seconds and the luminosity 
is $L_\gamma \propto E_{\gamma}^{-1/2}$ \cite{waxman1}. $E_{\gamma, min}$,
the minimum photon energy to produce pions via the
$\Delta$-resonance, is given by:
\begin{eqnarray}
E_{\gamma, min} = \frac{\gamma_{rs}^2(m_\Delta^2-m_p^2)}{4 E_{p}}.
\end{eqnarray}
Therefore $10^{21}$ eV protons can kinematically produce $\pi$'s on
photons with
energy as low
as 10 eV.  Combining Eq.\,66 and Eq.\,67 we find
\begin{eqnarray}
f_{\pi}\simeq .003 \times \bigg(\frac{10\, \rm{
eV}}{E_{\gamma,min}}\bigg) \bigg(\frac{E_p}{10^{21}\rm{eV}}\bigg)^{1/2}
\bigg(\frac{300}{\gamma_{rs}}\bigg)^5 \bigg(\frac{20 \rm{
sec}}{\Delta{t}}\bigg).
\end{eqnarray}
Note that above keV energy, the photon luminosity follows the broken
spectrum with $L_\gamma \propto E_\gamma^{-1}$ and, therefore,
$f_{\pi}\propto E_p$ rather than
$f_{\pi}\propto E_p^{1/2}$.

Associating the accelerated beam with the observed ultra high-energy
cosmic ray flux, $dN_p/dE_p \sim A E_p^{-2}$, where
$A\sim5 \times 10^{-4}\rm{m}^{-2}\rm{s}^{-1}\rm{sr}^{-1}\rm{GeV}$, and
using $E_\nu\sim .05 E_p$, see Eq.\,46, the resulting neutrino flux is
given by
\begin{eqnarray}
\frac{dN_\nu}{dE_\nu}(E_\nu) \sim \frac{dN_p}{dE_p}(E_p=20 E_\nu) \times
f_\pi(E_p=20 E_\nu),
\end{eqnarray}
\begin{eqnarray}
\frac{dN_\nu}{dE_\nu}(E_\nu) \sim A \times (20 E_\nu)^{-2}\times .003 
\times
\bigg(\frac{10 \rm{ eV}}{E_{\gamma,min}}\bigg) \bigg(\frac{20
E_\nu}{10^{21}\rm{eV}}\bigg)^{1/2}
\bigg(\frac{300}{\gamma_{\rm{rs}}}\bigg)^5 \bigg(\frac{20 \rm{
sec}}{\Delta{t}}\bigg),
\end{eqnarray}
\begin{eqnarray}
\frac{dN_\nu}{dE_\nu}E_\nu^2(E_\nu) \sim 2 \times 10^{-14}\,
E_\nu^{1/2}(\rm{GeV}) \times \bigg(\frac{10 \rm{
eV}}{E_{\gamma,min}}\bigg) \bigg(\frac{300}{\gamma_{\rm{rs}}}\bigg)^5
\bigg(\frac{20 \rm{ sec}}{\Delta{t}}\bigg) \rm{m}^{-2} \rm{s}^{-1}
\rm{sr}^{-1} \rm{GeV}.
\end{eqnarray}
It is important to note that if the burst occurs in a region of higher
density gas, as can be the case for a collapsing star,
reverse shocks are produced earlier and, therefore, with smaller
Lorentz factors.  This results in $f_\pi\simeq 1$.  Then,
\begin{eqnarray}
\frac{dN_\nu}{dE_\nu}(E_\nu) \simeq \frac{dN_p}{dE_p}(E_p=20 E_\nu)
\times
1 \simeq A \times (20 E_\nu)^{-2},
\end{eqnarray}
\begin{eqnarray}
\frac{dN_\nu}{dE_\nu}E_\nu^2(E_\nu) \sim 10^{-6} \rm{m}^{-2} \rm{s}^{-1}
\rm{sr}^{-1} \rm{GeV}.
\end{eqnarray}
In either case, the result is only valid above the threshold energy 
required to generates pions via the $\Delta$-resonance,
\begin{eqnarray}
E_\nu^{\rm{min}}\simeq .05 E_p^{\rm{min}}\simeq .05
\frac{\gamma^2_{rs}(m_\Delta^2-m_p^2)}{4 E_\gamma^{\rm{max}}}\sim 7\times
10^{17} \rm{eV}\times \bigg(\frac{\gamma_{rs}}{300}\bigg)^2 \bigg(\frac{1
\rm{keV}}{E_\gamma^{\rm{max}}}\bigg).
\end{eqnarray}
Below this threshold, ultra high-energy protons may still interact with
non-thermal MeV photons, however.

The event rate in a neutrino telescope is calculated following Eq.\,11. 
In the high-energy appoximation,
\begin{eqnarray}
P_{\nu \rightarrow \mu} \simeq  1.2 \times 10^{-2}
E^{0.5}_{\nu,obs}(\rm{EeV}).
\end{eqnarray}
This yields
\begin{eqnarray}
N_{events} \sim \int_{7\times 10^{8} \rm{GeV}}^{5 \times 10^{10}
\rm{GeV}}
10^{-6} E_\nu^{-2}\times 3.7 \times 10^{-7} E^{0.5}_{\nu}  dE_\nu \rm{
m}^{-2} \rm{s}^{-1} \rm{sr}^{-1}  \sim .01 \,\rm{ km}^{-2} \rm{yr}^{-1}.
\end{eqnarray}
This is a very small rate indeed. The neutrino energy is, however, above
the threshold for EeV telescopes using acoustic, radio or horizontal air
shower detection techniques. This mechanism may represent an opportunity
for detectors with very high threshold, but also large effective area to
do
GRB physics.

\subsubsection{The Decoupling of Neutrons: GeV Neutrinos}

The conversion of radiation into kinetic energy in the fireball will
accelerate neutrons along with protons, especially if the progenitor
involves neutron stars.  Protons and neutrons are initially coupled by
nuclear elastic scattering. If the expansion of the fireball is
sufficiently rapid the neutrons and protons will no longer interact.
Neutrons decouple from the fireball while protons are still accelerated.
Protons and neutrinos may then achieve relative velocities sufficient to 
generate pions which decay into GeV neutrinos \cite{gev1,gev2}.
We define the ratio of neutrons to protons as,
\begin{eqnarray}
\xi \equiv \frac{n_n}{n_p},
\end{eqnarray}
which initially remains constant during expansion.  The fraction of 
neutrons which generate pions is calculated in the same way as in 
Eq.\,47,
\begin{eqnarray}
f_{\pi}\simeq \frac{\Delta R^\prime}{\lambda_{pn}}.
\end{eqnarray}
We can relate the density of nucleons to the density of photons by the 
dimensionless entropy,
\begin{eqnarray}
n_{p+n}\simeq n_{\gamma}\frac{\gamma E_\gamma}{\eta m_p c^2}.
\end{eqnarray}
Following the arguments used in our discussion of PeV neutrinos, we arrive at
\begin{eqnarray}
f_{\pi} \simeq \frac{L_{\gamma}}{m_p \eta \gamma^3 \Delta t}
\frac{\sigma_{np}}{4 \pi c^2 (1+\xi)},
\end{eqnarray}
where $\sigma_{np}\simeq 3 \times 10^{26} \rm{cm}^2 \,$ is the 
neutron-proton cross section for pion production. As the neutrons and 
protons decouple, $f_{\pi}\,$ approaches unity. Using the fact that 
$\gamma$ asymptotically approaches $\eta$ at the end of expansion, we 
see that decoupling occurs for
\begin{eqnarray}
\eta \gsim \eta_{np} \equiv \bigg(\frac{L\sigma_{np}}{4 \pi R_0 m_p c^3
(\xi+1)}\bigg)^{1/4} \simeq 400\,
\bigg(\frac{L}{10^{52}\rm{erg/s}}\bigg)^{1/4}
\bigg(\frac{100\,\rm{km}}{R_0}\bigg)^{1/4}\bigg(\frac{2}{\xi+1}\bigg)^{1/4}
.
\end{eqnarray}
In fact, the requirement for exceeding the threshold for $\pi$ production
is $\eta \geq 1.2\, \eta_{np}$ \cite{gev1}. The scattering time is 
therefore longer than the expansion time by a factor $1.2^{4}\simeq 2.1$
and $1-e^{-2.1}\cong$ 88\% of the neutrons scatter.  If threshold were
exceeded by, say $\eta=1.5 \, \eta_{np}$, then more than 99\% of the
neutrons would scatter. It is therefore a reasonable approximation to
assume that all neutrons produce $\pi$'s as long as $\eta$ is above
threshold.
The number of neutrons in the fireball is large with
\begin{eqnarray}
N_n \simeq \frac{E}{m_p c^2} \frac{\xi}{1+\xi} \frac{1}{\eta} \sim 7
\times 10^{52} \bigg(\frac{E}{10^{53}
\rm{ergs}}\bigg)
\bigg(\frac{2\xi}{1+\xi}\bigg)\bigg(\frac{500}{\eta}\bigg).
\end{eqnarray}
Above the pion threshold, every neutron interacts with a proton producing one of the following:
\begin{itemize}

\item{}$p+n\rightarrow p+p+\pi^{-}\rightarrow \bar{\nu_{\mu}}+\mu^{-}\rightarrow e^{-}+\bar{\nu_{e}}+\nu_{\mu}+\bar{\nu_{\mu}}$

\item{}$p+n\rightarrow n+n+\pi^{+}\rightarrow \nu_{\mu}+\mu^{+}\rightarrow e^{+}+\nu_{e}+\bar{\nu_{\mu}}+\nu_{\mu}$

\item{}$p+n\rightarrow p+n+\pi^{0}\rightarrow \gamma+\gamma$

\end{itemize}
Thus on average, each interaction produces two 30-50 
MeV muon neutrinos and two 30-50 MeV electron neutrinos or two 70 MeV 
photons. The observed neutrino energy is
\begin{eqnarray}
E_{\nu, obs} \simeq 30-50\,\rm{MeV} \times \frac{\gamma}{1+z} \simeq
6-10\, GeV \times \bigg(\frac{\gamma}{400}\bigg)
\bigg(\frac{2}{1+z}\bigg).
\end{eqnarray}
This energy is below the threshold of neutrino telescopes with the
possible exception of Baikal and ANTARES provided it is built with a 
sufficiently
dense arrangement of the photomultipliers.
\begin{eqnarray}
P_{\nu \rightarrow \mu}(E_{\nu} \sim \rm{GeV}) \sim 10^{-7} 
E_{\nu,\rm{obs}}(GeV)
\end{eqnarray}
parametrizes the chance of detecting a $\sim$GeV muon neutrino in 
ANTARES as described by Eq.\,11. This leads to an event rate from a 
single burst of

\begin{eqnarray}
N_{events}\simeq N_{\nu}\, \frac{\rm{A}_{\rm{eff}}}{4\pi D^2} \, 
P_{\nu\rightarrow\mu}
\end{eqnarray}
where $N_{\nu}\,=2 N_n$ is the number of muon neutrinos emitted by the 
burst, $\rm{A}_{\rm{eff}}$ is the detector's effective area and $D$ is 
the distance to the burst. Positioning, for simplicity, all bursts at 
z=1, this reduces to

\begin{eqnarray}
N_{events}\sim N_{bursts} \times \bigg(2\, N_n\frac{A_{\rm{eff}}}{4\pi 
D^2} P_{\nu\rightarrow\mu}(8 \rm{GeV}) \bigg)
\end{eqnarray}
\begin{eqnarray}
\sim .1 \times \bigg(\frac{N_{\rm{bursts}}}{1000
\rm{yr}^{-1}}\bigg) \bigg(\frac{E}{10^{53} \rm{ergs}}\bigg) \bigg(\frac{2
\xi}{1+\xi}\bigg)   \bigg(\frac{3-2
\sqrt{2}}{2+z-2\sqrt{1+z}}\bigg) \bigg(\frac{A_{\rm{eff}}}{.1
\rm{km}^2}\bigg)^{3/2} \rm{yr}^{-1}.
\end{eqnarray}
This estimate is optimistic and the rate quite small. Observation
requires a burst with favorable fluctuations in distance and energy. We
discuss this important aspect of GRB detection next.

\subsubsection{Burst-To-Burst Fluctuations and Neutrino Event Rates}

We have focused so far on the diffuse flux of neutrinos produced
collectively by all GRB. Given the possibility to observe individual
GRB, it is
important to consider the fluctuations that may occur in the dynamics
from burst to burst. In the presence of large fluctuations, the relevant
observable becomes the number of neutrinos observed from individual
bursts \cite{fluc1,fluc2}. We will consider fluctuations in the distance
to the burst $D$, and in its energy $E$. One can further contemplate
fluctuations in the Lorentz factor, $\gamma$, or equivalently in $\eta$,
the dimensionless entropy, although it is not clear whether energy
and Lorentz factor are independent quantities.

The number of neutrinos from a single source at a distance $D$ is
proportional to $D^{-2}$. Considering burst-to-burst fluctuations in
distance will enhance the fluxes previously calculated. This can be see
as follows
\begin{eqnarray}
\frac{\int^{c t_{\rm{Hubble}}}_0 dN(D) D^{-2}}{\int^{c
t_{\rm{Hubble}}}_0 dN(D)D_{\rm{ave}}^{-2}}=\frac{\int^{c
t_{\rm{Hubble}}}_0 dD}{\int^{c t_{\rm{Hubble}}}_0 D^2
D_{\rm{ave}}^{-2} \rm{dD}}=\frac{D_{\rm{ave}}^{2} c
t_{\rm{Hubble}}}{\frac{1}{3}(c
t_{\rm{Hubble}})^3}=3 \frac{D_{\rm{ave}}^{2}}{(c t_{\rm{Hubble}})^2},
\end{eqnarray}
where we have assumed an isotropic Euclidian distribution of sources. In
contrast, the average distance of a burst $D_{\rm{ave}}$  is
\begin{eqnarray}
\int^{D_{\rm{ave}}}_0 D^2 dD =\frac{1}{2} \int^{c t_{\rm{Hubble}}}_0 D^2
dD \rightarrow D_{\rm{ave}}=2^{-1/3} c t_{\rm{Hubble}}.
\end{eqnarray}
Therefore, the event rate is enhanced by spatial fluctuations by a factor
\begin{eqnarray}
\rm3\times 2^{-2/3} \simeq 1.9.
\end{eqnarray}
This enhancement factor should be applied to the diffuse flux. It also
represents an increased probability to observe a single burst
yielding multiple neutrino events. Assuming a more realistic
cosmological distribution for starburst galaxies increases the effect
of spatial fluctuations by an additional factor of 2 above the Euclidian
result.

Next, we consider energy. Observations show that approximately one out
of ten bursts is ten times more energetic than the median burst and
approximately one out of a hundred is
a hundred times more energetic.  This results in an enhancement of the
neutrino signal by a factor of
\begin{eqnarray}
{.89\times E_{\rm{median}} + .10 \times
10\,E_{\rm{median}} + .01 \times 100\,E_{\rm{median}}}\simeq
2.9 \, {E_{\rm{median}}}.
\end{eqnarray}
If instead of a step function, a smooth distribution
$\frac{dN}{dE}\propto E^{-2}$ covering two orders of magnitude is used,
the enhancement factor is increased to 4.6.

We thus reach the very important conclusion that correct averaging of
bursts over their cosmological spatial distribution and over their
observed energy distribution enhances the neutrino signal by
approximately one order of magnitude. Variations in distance
and energy affect all GRB neutrino fluxes previously discussed,
regardless of the production mechanism.

The third, and most important, quantity to vary from burst-to-burst is
the Lorentz factor, or entropy. Its variation can greatly modify some
fluxes, e.g. as
$\sim\gamma^4$ for PeV neutrinos or $\sim \gamma^5$ for EeV
neutrinos, and it barely affects others, e.g. GeV neutrinos produced by
decoupled neutrons. The range over which $\gamma$ can vary is, however,
limited \cite{waxman1,waxman2,halzenlec,piran}. The Lorentz
factor  i) must be roughly $300$ to generate the observed
non-thermal gamma ray spectrum, ii) must be large enough to produce the
highest energy cosmic rays of $\sim 10^{20}$eV, and iii) the energy of
shocked protons must be sufficient to be above threshold for producing
pions on fireball gamma rays.

The last condition requires that
\begin{eqnarray}
m_p \gamma \gsim \frac{m_\Delta^2-m_p^2}{4 E_{\gamma,\rm{obs}}}
\Longrightarrow \gamma \gsim 170 \bigg(\frac{1
\,\rm{MeV}}{E_{\gamma,\rm{obs}}}\bigg),
\end{eqnarray}
and the second that
\begin{eqnarray}
\gamma \gsim A^{1/2}\times 130 \bigg(\frac{E}{10^{20}\rm{eV}}\bigg)^{3/4}
\bigg(\frac{.01 \rm{sec}}{\Delta t}\bigg)^{1/4}
\end{eqnarray}
following Eq.\,37.  Once again, A is a factor of order unity, $E$ the
maximum proton energy generated and $\Delta t$ the time scale
associated with the rapid variations observed in single bursts.

Together, these constraints limit any variation of the Lorentz factor to
at most one order of magnitude.  This does, however,  correspond to a
variation over four (five) orders of magnitude in the fraction of energy
converted to
$\pi$'s which yield PeV (EeV) neutrinos. A change in Lorentz factor
modifies the peak neutrino energy of the neutrinos as
$\gamma^2$ and is, therefore, a secondary effect relative to
the fraction of energy converted to pions.

How the Lorentz factor of GRB vary from burst-to-burst is an open
question. It is interesting to note that even with conservative
fluctuations of a factor of two or so (say between 200 and 400), the
difference in neutrino flux exceeds one order of
magnitude.  Combined with variations in distance, and perhaps burst
energy, the occasional nearby and bright (small $\gamma$, high $E$, or
both) burst becomes a superior experimental signature to the diffuse
flux. More detailed modeling has been presented in Ref.\,\cite{fluc1,fluc2}. For an alternative perspective, see reference \cite{waxman11}.

\subsubsection{The Effect of Neutrino Oscillations}

Perhaps the most important discovery of particle physics in the last decade is the oscillation of neutrinos. The fact that neutrinos can change flavor as they propagate can have an effect on the neutrino fluxes observed on earth. For propagation over cosmological distances, the neutrino survival probablities become simple. For the example of neutrinos from pion decay, where there is an initial ratio of 1 to 2 to 0 for electron, muon and tau flavors, respectively, we expect approximately a ratio of 1 to 1 to 1 after oscillations. Therefore most of the event rates calculated for muon neutrinos have been overestimated by a factor of two. At sufficient energies, however, electron and tau neutrinos can be observed as well thus counteracting this effect in part.  

For a review of neutrino oscillations, see Ref.\,\cite{osc}.

\subsection{Blazars: the Sources of the Highest Energy Gamma rays}

\subsubsection{Blazar Characteristics}

Active Galactic Nuclei (AGN) are the brightest sources in the universe.
They are of special interest here because some emit most of their 
luminosity at
GeV energy and above. A subset, called blazers, emit high-energy
radiation in collimated jets pointing at the earth. They have the
following characteristics:

\begin{itemize}

\item Although less luminous than GRB, with inferred isotropic
luminosities of $\sim10^{45}-10^{49}\rm{erg}\/\rm{s}$, they radiate this
luminosity over much longer time periods with regular flares extending
for days \cite{redagn1,histagn3,sikora}. The energetics require a black
hole
roughly one billion times more massive than our sun.

\item Blazars produce radiation from radio waves to TeV gamma rays with
enhancements in $E^2 dN/dE$ or $\nu F(\nu)$ in the IR to X-ray and the
MeV-TeV range \cite{spec1,spec2,spec3,spec4,AGN1}.  Roughly 60
sources have been observed in the MeV-GeV range by the EGRET instrument
on the Compton Gamma ray Observatory. A handful of TeV observations have
been reported thus far \cite{weekes,tevagn2,redagn1,
highest,prot,egret,egret2}.  EGRET has found that blazers produce a
typical $\phi_\gamma
\propto E^{-2.2}$ spectrum in the
MeV-GeV range.  This spectrum may extend above or below this
range \cite{redagn1, histagn2, sikora}.  There appears to be an inverse
relationship between blazar luminosity and peak emission.  High
luminosity
blazars tend to peak in the GeV and optical bands 
\cite{sikora,spec3,high,high2}, while low luminosity blazars tend to
peak
in the TeV and X-ray bands \cite{sikora,spec1,low1,low2,low3,high2}.

\item Blazars are cosmological sources. They appear less distant than
GRB only because their lower luminosity makes distant sources difficult 
to
observe \cite{redagn1,sikora}.

\item The time scale of flaring in blazar luminosity varies from a
fraction of a day to years.  This range of time scales indicates a
sub-parsec engine; $c \Delta t \sim R^{\prime} \rightarrow R^{\prime}
\sim 10^{-3}-10^{-1}$
pc \cite{redagn1,histagn2,histagn22, sikora}.

\end{itemize}

Prior to the launch of EGRET, the only survey above 100 MeV energy had
been made by
the COS-B satellite.  The COS-B survey revealed the first extragalactic 
gamma ray source, the active galaxy 3C-272 \cite{histagn1}. After EGRET 
was launched in 1991, several gamma ray
blazars were
discovered. The majority of these were flat-spectrum radio-loud quasars, 
although some were BL Lac objects which emit into the TeV
range \cite{histagn2,histagn22, lep1}.

Three of the EGRET sources have also been observed by ground-based
atmospheric
Cerenkov telescopes previously discussed.  Future observations of 
blazars will include
lower-threshold (below 50 GeV) ground-based telescopes, as well as next 
generation satellites
such as GLAST \cite{glast, histagn3,redagn1}.

\subsubsection{Blazar Models}

It is widely believed that blazars are powered by accreting supermassive
black holes with masses of $\sim 10^7 M_{\odot}$ or more. Some of the
infalling matter is reemitted and accelerated in highly beamed jets
aligned with the rotation axis of the black hole.

It is generally agreed upon that synchrotron radiation by accelerated
electrons is the source of the observed IR to X-ray peak in the
spectrum \cite{lep1,lep2,lep3,lep4,sikora}.  Inverse Compton scattering
of synchrotron or, possibly, other ambient photons by the same electrons
can accommodate all observations
of the MeV-GeV second peak in the spectrum. There is a competing
explanation for the second peak, however. In hadronic models \cite{had1,
had2,had3,had4,had5,had6,had7,sikora}, MeV-GeV gamma rays are generated
by
accelerated protons interacting with gas or
radiation surrounding the black hole. Pions produced in these 
interactions decay into the
observed gamma rays and not-yet observed neutrinos.  This process is 
accompanied by synchrotron radiation of the protons.

\begin{figure}[thb]
\centering\leavevmode
\includegraphics[width=5in]{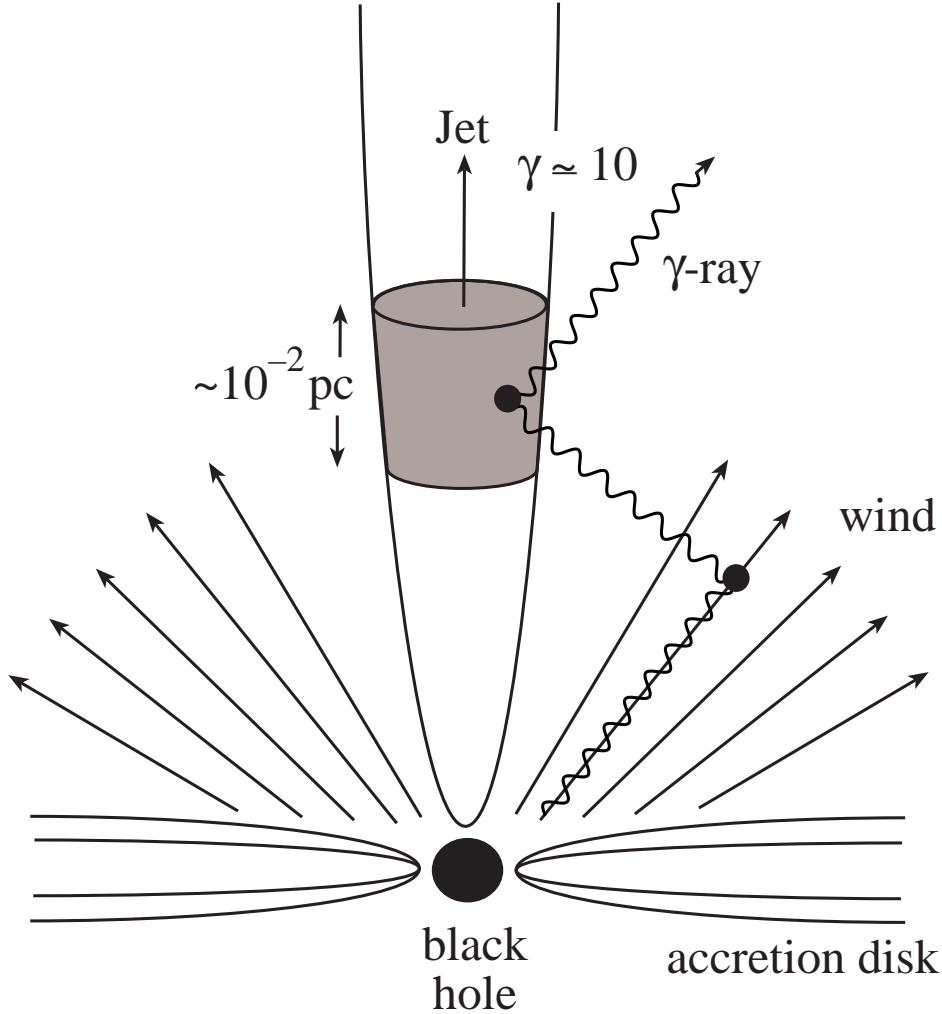}
\caption{Diagram of Blazar Kinematics}
\label{eighteen}
\end{figure}

The basic dynamics of the blazar is common to all models.  Relativistic
jets are generated with substructure that takes the form of ``blobs''
or ``sheets'' of matter traveling along the jet with Lorentz factors of
10-100.  As previously discussed, in order to accommodate the
observation of flares, the thickness of these sheets must be less than
$\gamma c \Delta t \sim 10^{-2}$ parsec, much smaller than their
width, which is of the order of 1 parsec.  It is in these blobs that
shocks produce TeV
gamma rays and high-energy neutrinos.

Several calculations of the neutrino flux from active galactic nuclei
have been performed 
\cite{agnnu1,agnnu2,agnnu3,agnnu4,agnnu5,agnnu6,halzenlec,berez10,dermer1,
dermer2,
external, external2, pohl}.  In the following sections, we describe two
examples that illustrate the mechanisms for neutrino production in
hadronic
blazars.

\subsubsection{Highly Shocked Protons: EeV Blazar Neutrinos}

If protons are present in blazar jets, they may interact with photons
via the $\Delta$-resonance to generate pions which then decay into very
high-energy neutrinos.  This process is similar to the process
generating PeV and EeV neutrinos in GRB.  There are
some important differences, however.  First, the Lorentz factor of the 
motion of
the blob, traveling towards the observer, is typically smaller than for 
GRB shells.
It can be constrained by considering the energy carried by the highest
energy gamma rays observed
in blazars.  These gamma rays escape the blob and must, therefore, be
below the
energy threshold for pair production with ambient
photons whose energy typically peaks
around $\sim 10\,$ eV; the UV bump.  Contemplating the observations of
gamma rays
above 15 TeV in Markarian 501 \cite{prot, highest}, evading pair
production requires
\begin{eqnarray}
E_{\gamma \rm{, max}} E_{\gamma \rm{, peak}} < \gamma^2 m_e^2,
\end{eqnarray}
\begin{eqnarray}
\gamma > 25 \bigg(\frac{E_{\gamma \rm{, max}}}{15\, \rm{TeV}}\bigg)^{1/2}
\bigg(\frac{E_{\gamma \rm{, peak}}}{10\, \rm{eV}}\bigg)^{1/2}.
\end{eqnarray}
We will therefore consider Lorentz factors in the range of 10 to 100.

A second difference between blazars and GRB is the geometry of the
shocked
material.  Instead of a shell, blobs can be treated as roughly 
spherical.  The energy
density is
\begin{eqnarray}
U^\prime_\gamma = \frac{L^\prime_\gamma \Delta t }{\frac{4}{3}\pi
R^{\prime 3} } = \frac{L_\gamma \Delta t }{\gamma \frac{4}{3}\pi
(\gamma c
\Delta t)^3 } = \frac{3 L_\gamma}{4 \pi c^3 \gamma^4 \Delta t^2}
\end{eqnarray}
Except for geometry, this is identical to Eq.\,49. We obtain a number
density of photons
\begin{eqnarray}
n_\gamma =\frac{U^\prime_\gamma}{E^\prime_\gamma} = \frac{3 L_\gamma}{4
E_\gamma \pi c^3 \gamma^3 \Delta t^2}
\end{eqnarray}
Following the arguments leading to Eqs.\,51, we obtain a rather large
conversion
of energy into pions
\begin{eqnarray}
f_{\pi} \simeq \frac{R^\prime}{\lambda_{p \gamma}} \simeq
\frac{L_{\gamma}}{E_{\gamma}}\frac{1}{\gamma^2 \Delta t} \frac{3
\sigma_\Delta \langle x_{p \rightarrow \pi} \rangle}{4\pi c^2} \simeq .35
\times \bigg(\frac{L_\gamma}{10^{45}\rm{erg/s}}\bigg) \bigg(\frac{10\,
\rm{eV}}{E_\gamma}\bigg)  \bigg(\frac{30}{\gamma}\bigg)^2
\bigg(\frac{1000\, \rm{sec}}{\Delta{t}}\bigg)
\end{eqnarray}
When $f_\pi$ approaches unity, pions will be absorbed before decaying
into neutrinos, requiring the substitution of $f_\pi$ by $1-e^{-f_\pi}$.
Note that fluctuations in $\gamma$ are not as important as for GRB
because the fraction of energy transferred to pions varies as
$\gamma^2$ rather than $\gamma^4$.  Moreover, it quickly saturates near
unity, especially for relatively low $\gamma$, i.e. bright
flares or short time scales.

For protons to photoproduce pions on photons with the ubiquitous UV 
photons of $\sim 10
\,$eV energy,
\begin{eqnarray}
E^\prime_p > \frac{m_\Delta^2-m_p^2}{4 E^\prime_\gamma}.
\end{eqnarray}
Therefore, in the observer's frame,
\begin{eqnarray}
E_p > 1.4 \times 10^{19} \rm{eV} \bigg(\frac{\gamma}{30}\bigg)^2
\bigg(\frac{10\, eV}{E_\gamma}\bigg).
\end{eqnarray}
If blazars are the sources of the highest energy cosmic rays, protons
are accelerated to this energy and will generate accompanying neutrinos
with energy
\begin{eqnarray}
E_\nu = \frac{1}{4} \langle x_{p \rightarrow \pi} \rangle  E_p  > 7
\times
10^{17} \rm{eV} \bigg(\frac{\gamma}{30}\bigg)^2 \bigg(\frac{10\,
\rm{eV}}{E_\gamma}\bigg).
\end{eqnarray}
The neutrino flux from blazars can be calculated in the same way as
for GRB
\begin{eqnarray}
\phi_\nu \simeq \frac{c}{4 \pi}\frac{1}{E_\nu}\bigg( \frac{1}{2}
(1-e^{-f_\pi})\, t_H \frac{dE}{dt} \bigg) e^{(1-e^{-f_\pi})},
\end{eqnarray}
\begin{eqnarray}
\phi_\nu = 10^{-15} \rm{cm}^{-2} \rm{s}^{-1} \bigg( \frac{7 \times
10^{17}\rm{eV}}{E_\nu} \bigg) \bigg(\frac{t_H}{10 \rm{Gyr}} \bigg) \bigg(
\frac{dE/dt}{4 \times 10^{44}} \bigg) (1-e^{-f_\pi})e^{(1-e^{-f_\pi})},
\end{eqnarray}
using Eq.\,52 and Eq.\,53. This a flux on the order of $100\, \rm{km}^{-2}
\rm{yr}^{-1}$ over $2\pi$ steridian. The number of detected events is
obtained from Eqs.\,11,
\begin{eqnarray}
N_{\rm{events}} \sim \phi_\nu  P_{\nu \rightarrow \mu} \sim 10 \,
\rm{km}^{-2} \rm{yr}^{-1} \bigg( \frac{7 \times 10^{17}\rm{eV}}{E_\nu}
\bigg)^{1/2} \bigg(\frac{t_H}{10 \rm{Gyr}} \bigg) \bigg( \frac{dE/dt}{4
\times 10^{44} \rm{Mpc}^{-3} \rm{yr}^{-1}} \bigg)
\end{eqnarray}
for $f_\pi\simeq.35$.  For values of $\gamma$ varying from 10 to 100, the
number of events varies from 1 to 70 events $\,\rm{km}^{-2}
\rm{yr}^{-1}$,
respectively. Observation in a kilometer-scale detector should be
possible.

The greatest uncertainty in this calculation is associated with the
requirement that
blazars accelerate protons to the highest observed energies.  In the 
next section we present an alternative mechanism
for producing neutrinos in blazars which does not invoke protons of such
high-energy.

\subsubsection{Moderately Shocked Protons: TeV Blazar Neutrinos}

In line-emitting blazers, external photons with energies in the
keV-MeV energy range are known to exist. They are clustered in clouds of
quasi-isotropic radiation. Protons of lower energy relative to those
contemplated in the previous section, can photoproduce pions in
interactions with these clouds \cite{dermer1,dermer2, external,
external2}. Consider a target of external photons of energy near
$E_{\gamma,\rm{ext}}$ with a luminosity $L_{\gamma,\rm{ext}}$. The
fraction of proton energy transfered to pions is approximately given by
\begin{eqnarray}
f_{\pi, \rm{ext}} \simeq f_{\pi, \rm{int}}
\frac{L_{\gamma,\rm{ext}}}{L_{\gamma,\rm{int}}}
\frac{E_{\gamma,\rm{ext}}}{E_{\gamma,\rm{int}}}.
\end{eqnarray}
The neutrino energy threshold is
\begin{eqnarray}
E_\nu > 7 \times 10^{13} \rm{eV} \bigg(\frac{\gamma}{30}\bigg)^2
\bigg(\frac{100\, \rm{keV}}{E_{\gamma,\rm{ext}}}\bigg).
\end{eqnarray}
We will no longer relate the flux of protons to cosmic rays.  Instead we
introduce the luminosity of protons, $L_p$ above pion production
threshold.  This is a largely unknown parameter although it has been
estimated to be on the order of 10\% of the total
luminosity \cite{berez10}.  The proton energy needed to exceed
the threshold of Eq.\,44 is
\begin{eqnarray}
E_p > 1.4 \times 10^{15} \bigg(\frac{\gamma}{30}\bigg)^2 \bigg(\frac{100
\, \rm{keV}}{E_{\gamma,\rm{ext}}}\bigg).
\end{eqnarray}
The neutrino flux can be calculated as a function of $L_p$
\begin{eqnarray}
\Phi_\nu \simeq \frac{  \frac{1}{2} \langle x_{p \rightarrow \pi} \rangle
L_p f_{\pi, \rm{ext}} \Delta t   }{ E_\nu 4 \pi D^2},
\end{eqnarray}
where $\Delta t$ is the duration of a blazar flare.  This reduces to
\begin{eqnarray}
\Phi_\nu \sim 4 \times 10^4 \,{\rm{km}}^{-2} \bigg(\frac{f_{\pi,
\rm{ext}}}{.5}\bigg) \bigg( \frac{L_p}{10^{45}\,\rm{erg/s}}\bigg) \bigg(
\frac{\Delta
t}{1000\,\rm{sec}}\bigg) \bigg(\frac{1000\,\rm{Mpc}}{D}\bigg)^2
\bigg(\frac{30}{\gamma} \bigg)^2
\bigg(\frac{E_{\gamma,\rm{ext}}}{100\,\rm{keV}}\bigg)
\end{eqnarray}
for a fifteen minute flare. Using Eqs.\,11, the event rate of TeV
neutrinos is
\begin{eqnarray}
N_{\rm{events}} \sim \phi_\nu  P_{\nu \rightarrow \mu} \sim 2 \,
{\rm{km}}^{-2}  \bigg(\frac{f_{\pi, \rm{ext}}}{.5}\bigg) \bigg(
\frac{L_p}{10^{45}\,\rm{erg/s}}\bigg) \bigg( \frac{\Delta
t}{1000\,\rm{sec}}\bigg) \bigg(\frac{1000\,\rm{Mpc}}{D}\bigg)^2
\bigg(\frac{30}{\gamma} \bigg)^{2/5}
\bigg(\frac{E_{\gamma,\rm{ext}}}{100\,\rm{keV}}\bigg)^{1/5}.
\end{eqnarray}
Note that this result is for a typical, but fairly distant source.  A
nearby line-emitting blazar could be a
strong candidate for neutrino observation. Considering the more than 60
blazars which have been observed, the total
flux may generate conservatively  tens or, optimistically, hundreds of
TeV-PeV neutrino events per year in a kilometer scale neutrino telescope
such as IceCube.  The upper range of this estimate can be explored by
the AMANDA experiment.

Blazar neutrino searches should be able to find incontrovertible
evidence for cosmic ray acceleration in active galaxies, or,
alternatively, challenge the possibility that AGN are the sources of the
highest energy cosmic rays.

\subsection{Neutrinos Associated With Cosmic Rays of Top-Down Origin}

In addition to astrophysical objects such as GRB and blazars, a variety
of top-down models have been proposed as the source of the highest
energy cosmic rays.  For example, annihilating or decaying superheavy
relic particles could produce the highest energy cosmic rays 
\cite{wim1,wim,wim2,wim3,wim4,wim5,wim6,wim6b,sh1,sh4}.  In addition to the
cosmic ray nucleons, they will also generate gamma-rays and neutrinos.
Other top-down scenarios which solve the cosmic ray problem in a similar
way include topological defects \cite{top1,top2,top3,top4,top5,top6} and Z-bursts \cite{z1,z2,z3,z4,z5}.
Conventional particle physics implies that ultra high-energy jets
fragment predominantly into photons with a small admixture of protons 
\cite{lepc1,lepc2,lepc3}.  This seems to be in disagreement with
mounting evidence that the highest energy cosmic rays are not photons 
\cite{WatsonZas}. In light of this information, we must assume that
protons, and not gamma-rays, dominate the highest energy cosmic ray
spectrum.  This does not necessarily rule out superheavy particles as
the source of the highest energy cosmic rays.  The uncertainties
associated with the universal radio background and the strength of
intergalactic magnetic fields leave open the possibility that ultra 
high-energy photons may be depleted from the cosmic ray spectrum near
$10^{20}$ eV, leaving a dominant proton component at GZK energies 
\cite{radio1,radio2,radio3,radio4}.  With this in mind, one must
normalize the proton spectrum from top-down scenarios with the observed
ultra high-energy cosmic ray flux.

An important point is that this ``renormalization'' is not only
challenged by the large sub-GeV photon flux, but also by the neutrino
flux associated with these models.  Neutrinos, which are produced more
numerously than protons
and travel much greater distances, typically provide an observable
signal in operating high-energy neutrino telescopes.

\subsubsection{Nucleons in Top-Down Scenarios}

The assumption that nucleons from ultra high-energy fragmentation are the source of the highest energy cosmic rays normalizes the rate of their generation.  To do this, it is
necessary to calculate the
spectrum of nucleons produced in such jets.  Each
jet will fragment into a large number of hadrons.  The quark
fragmentation function can be parametrized as \cite{frag}:
\begin{eqnarray}
\frac{dN_{\rm{H}}}{dx}= C x^{-3/2}(1-x)^2.
\end{eqnarray}
Here, $x=E_{\rm{Hadron}}/E_{\rm{Jet}}$, $N_{\rm{H}}$ is the number of
hadrons and $C=15/16$ is a normalization constant determined by energy
conservation.  For a more rigorous treatment of fragmentation, see 
\cite{sh2,sh3,drees}.

At the energies considered, all flavors of quarks
are produced equally.  Top quarks immediately decay into $bW^{\pm}$
pairs.  Bottom and charm quarks lose energy from
hadronization before decaying into charmed and other hadrons.  Hadrons
eventually decay into pions and nucleons.  The injection
spectrum of nucleons produced in parton jets can be approximately
described by \cite{specnuc}:
\begin{eqnarray}
\Phi_N(E) \simeq \frac{dn_X}{dt} N_q^2 \frac{f_N}{E_{\rm{Jet}}}
\frac{dN_{\rm{H}}}{dx}.
\end{eqnarray}
Here, $\frac{dn_X}{dt}$ is the number of jets produced per
second per cubic meter.  $N_q$ is the number of quarks produced per
jet in the energy range concerned.  $f_N \sim .03$ is the nucleon
fraction in the jet from a single quark \cite{specnuc}.

To solve the ultra high-energy cosmic ray problem, this proton flux
must accommodate the events above the GZK cutoff.  Observations
indicate on the order of $10^{-27}$ events $\rm{m}^{-2} \rm{s}^{-1}
\rm{sr}^{-1} \rm{GeV}^{-1}$ in the energy range above the GZK cutoff ($5
\times 10^{19}$ eV to $2 \times 10^{20}$ eV) \cite{flyes,agasa}. The
formalism of a generic top-down scenario is sufficiently flexible to
explain the data from either the HIRES or AGASA experiments. The distribution of ultra high-energy jets can play an important role in the spectra of nucleons near the GZK cutoff. For example, the distribution for decaying or annihilating dark matter is likely to be dominated by the dark matter within our galaxy. This overdensity strongly degrades the effect of the GZK cutoff.

\subsubsection{Neutrinos in Top-Down Scenarios}

There are several ways neutrinos can be produced in the fragmentation of
ultra high-energy jets.  First bottom and charm quarks decay
semileptonically about 10\% of the time.  Secondly, the cascades
of hadrons produce mostly pions.  About two thirds of these
pions will be charged and decay into neutrinos \cite{specnuc}.
Furthermore, top quarks produced in the jets decay nearly 100\% of the
time to $bW^{\pm}$.  The $W$ bosons then decay semileptonically
approximately 10\% of the time to each neutrino species.

Generally, the greatest contribution to the neutrino spectrum is from
charged pions. The injection spectrum of charged pions is given
by \cite{specnuc}:
\begin{eqnarray}
\Phi_{(\pi^+ + \pi^-)}(E) \simeq \frac{2}{3}
\frac{(1-f_N)}{f_N}\Phi_N(E) 
\end{eqnarray}
\begin{eqnarray}
\Phi_{(\pi^+ + \pi^-)}(E) \simeq \frac{2}{3}
(1-f_N)\frac{dn_{X}}{dt}
\frac{N_q^2}{E_{\rm{Jet}}}\frac{dN_{\rm{H}}}{dx}.
\end{eqnarray}
The resulting injection of neutrinos is given by \cite{specsun}:
\begin{eqnarray}
\Phi_{(\nu + \bar{\nu})}(E) \simeq 2.34 \int_{2.34 E}^{E_{Jet}/N_q}
\frac{dE_{\pi}}{E_{\pi}} \Phi_{(\pi^+ + \pi^-)}(E_{\pi}).
\end{eqnarray}
Using $f_N \simeq .03$ and $\frac{dn_X}{dt} \simeq 1.5
\times 10^{-37}$, this becomes:
\begin{eqnarray}
\Phi_{(\nu + \bar{\nu})}(E) \sim 3.0\times 10^{-36}   \int_{2.34
E}^{E_{Jet}/N_q} \frac{dE_{\pi}}{E_{\pi}} N_q^2
(1-\frac{E_{\pi}}{E_{\rm{Jet}}})^2
\frac{E_{\rm{Jet}}^{.5}}{E_{\pi}^{1.5}}
\end{eqnarray}
for each species of neutrino.  $N_q$ is the number of quarks produced
in the fragmentation in the energy range of interest.

To obtain the neutrino flux, we multiply the injection spectrum by the
average distance traveled by a neutrino and by the rate per volume for
hadronic jets which we calculated earlier.  Neutrinos, not being
limited by scattering, travel up to the age of the universe at the
speed of light ($\sim$ 3000 Mpc in an Euclidean approximation).  A
random cosmological distribution of ultra high-energy jets provides an
average distance between 2000 and 2500 Mpc.

The neutrinos generated in these scenarios can be constrained by measurements of the  high-energy diffuse flux. AMANDA-B10, with an effective area of
$\sim$5,000 square meters has placed the strongest limits on the flux at
this time. In addition to the diffuse flux of high-energy neutrinos, the number of extremely high-energy events can be considered. Depending on the details of fragmention and jet distribution, tens to thousands of events per year per
square kilometer effective area can be generated above an energy threshold of 1 PeV where there are no significant backgrounds to interfere with the signal.

As a simple example, take the Z-burst scenario. In this scenario, ultra high-energy neutrinos travel cosmological distances and interact with massive ($\sim$eV) cosmic background neutrinos at the Z-resonance. The Z bosons then decay producing, among other things, the super-GZK cosmic rays.  In the center-of-mass frame of the neutrino 
annihilation the $Z$ is produced at rest with all the features of its decay 
experimentally known. Independent of any differences in the calculation, 
normalizing the cosmic ray flux to the protons, rather than the photon 
flux, raises the sensitivity of neutrino experiments as in all other 
examples. This can be demonstrated with a simple calculation. Data 
determine that Z-decays produce 8.7 charged pions for every proton and, 
therefore, $8.7\times 3 = 26.1\,$ neutrinos from 
$\pi^{\pm}\rightarrow \bar{\nu_{\mu}} \mu \rightarrow e \nu_{\mu} \bar{\nu_e}$ for every proton.  AGASA data, with an 
integrated proton flux of $\sim 5 \times 10^{24}\, \rm{ev}^2\, 
\rm{m}^{-2}\, \rm{s}^{-1}\, \rm{sr}^{-1}\,$ in the range of, say, 
$3\times 10^{19}\, \rm{to}\, 2 \times 10^{20} \rm{eV}$, indicate $\sim 
.5\,$ protons per square kilometer, per year over $2\pi\,$ steridians. 
This normalization, corrected for the fact that neutrinos travel 
cosmological distances rather than a GZK radius for protons, predicts 
$.5 \times 26.1 \times \frac{3000\, \rm{Mpc}}{50\, \rm{Mpc}} \sim 
800 \,$ neutrinos per square kilometer, per year over $2\pi\,$ 
steridians. We here assumed an isotropic distribution of cosmological 
sources. The probability of detecting a neutrino at $\sim 10 \,$ EeV is 
$\sim .05$, see Eq.\,11. Therefore, we expect 40 events per year 
in IceCube, or a few events per year in AMANDA II. As with other top-down 
scenarios, present experiments are near excluding or 
confirming the model.

Using the high-energy neutrino diffuse flux measurements and searches
for super-PeV neutrinos, top-down scenarios can be constrained.
Further data from AMANDA, or next generation neutrino telescope IceCube,
will test the viability of top-down scenarios which generate the highest
energy cosmic rays.

\section{The Future for High-energy Neutrino Astronomy}

At this time, neutrino astronomy is in its infancy. Two telescopes, 
one in Lake Baikal and another embedded in the South Pole glacier, 
represent proof of concept that natural water and ice can be transformed 
into large volume Cherenkov detectors. With an acceptance of order 
$0.1\,km^2$, the operating AMANDA II telescope represents a 
first-generation instrument with the potential to detect neutrinos from 
sources beyond the earth's atmosphere and the sun. It has been operating 
for 3 years with 302 OM and for almost 3 years with 677 OM. Only 1997 data 
have been published.
While looking forward to AMANDA data, construction has started on 
ANTARES, NESTOR and IceCube, with first deployments anticipated in 2002 
and 2003. At super-EeV energies these experiments will be joined by HiRes, 
Auger and RICE. A variety of novel ideas exploiting acoustic and radio 
detection techniques are under investigation, including ANITA for which 
a proposal has been submitted. Finally, initial funding of the R\&D 
efforts towards the construction of a kilometer-scale telescope in the 
Mediterranean has been awarded to the NEMO collaboration.
With the pioneering papers published nearly half a century ago by 
Greisen, Reines and Markov, the technology is finally in place for 
neutrino astronomy to become a reality. The neutrino, a particle that is 
almost nothing, may tell us a great deal about the universe \cite{sutton}.

\begin{acknowledgments}
This work was supported in part by a DOE grant No. DE-FG02-95ER40896 and 
in part by the
Wisconsin Alumni Research Foundation.
\end{acknowledgments}

\end{document}